\documentclass[final]{siamltex}
\usepackage{amsmath,amssymb,epsfig}
\def\theequation{\thesection.\arabic{equation}}

\newtheorem{cor}[theorem]{Corollary}

\newcommand{\eqa}{\begin{eqnarray}}
\newcommand{\eeqa}{\end{eqnarray}}
\newcommand{\beq}{\begin{equation}}
\newcommand{\eeq}{\end{equation}}
\newcommand{\nn}{\nonumber}
\newcommand{\epf}{$\quad$\hfill
\raisebox{0.11truecm}{\fbox{}}\par\vskip0.4truecm}
\newcommand{\pf}{\noindent{\it Proof \ }}
\newcommand{\pal}{\partial}

\begin{document}
\title{Numerical Study of breakup in generalized Korteweg-de Vries 
and Kawahara equations}

\author{B.~Dubrovin\thanks{SISSA, Via Bonomea 265, I-34136 Trieste, Italy, {\tt 
dubrovin@sissa.it} and Laboratory of Geometric Methods in Mathematical 
Physics, Moscow State University `M.V.Lomonosov',}
			 \and
T.~Grava\thanks{SISSA, Via Bonomea 265, I-34136 Trieste, Italy, {\tt 
grava@sissa.it}} \and
C.~Klein\thanks{Institut de Math\'ematiques de Bourgogne,
		Universit\'e de Bourgogne, 9 avenue Alain Savary, 21078 Dijon
		Cedex, France, 
    {\tt christian.klein@u-bourgogne.fr}}}

\maketitle

\begin{abstract}
This article is concerned with a conjecture in \cite{dub06} on the formation of 
dispersive shocks in a class of Hamiltonian  dispersive regularizations of the 
quasilinear transport equation. The regularizations are characterized 
by two arbitrary functions of one variable, where the
 condition of integrability implies   that one of these functions  must not vanish. 

It is shown numerically for a large class of equations that the local 
behaviour of their solution near the point of gradient catastrophe 
for the transport equation is described locally by a special solution 
of a Painlev\'e-type equation. This local description holds also for solutions to equations where blow up can occur in finite time. 

Furthermore, it is shown that a solution of the dispersive 
equations  away from the point of gradient 
catastrophe is approximated by a solution of the transport equation  with the same initial data, 
modulo terms of order $\epsilon^2$  where $\epsilon^2$ is the  small 
dispersion parameter. Corrections up to order $\epsilon^4 $ are obtained and tested numerically.

\end{abstract}

\begin{keywords}
Generalized Korteweg-de Vries equations, Kawahara equations, 
dispersive shocks, multi-scales analysis
\end{keywords}

\begin{AMS}
    Primary, 65M70; Secondary, 65L05, 65M20
\end{AMS}
\section{Introduction}
Many wave phenomena in dispersive media with negligible dissipation, 
in hydrodynamics, nonlinear optics, and plasma physics are described by
nonlinear dispersive partial differential equations (PDE). These 
equations are also
mathematically challenging since the solutions can have highly 
oscillatory regions and  blowup even for smooth initial data (see, e.g., \cite{zk},  \cite{gp}, \cite{hl}). 

This article is concerned with a conjecture in \cite{dub06} on the formation of 
dispersive shocks \cite{gp}, \cite{ks}, \cite{sul}  in a class of Hamiltonian regularizations of the 
quasilinear transport equation
\beq\label{tr1}
u_t+a(u)u_x=0, \quad a'(u)\neq 0,\quad u,x\in\mathbb{R}.
\eeq
In the present paper we will consider general Hamiltonian perturbations of \eqref{tr1} up to fourth order in a small dispersion parameter 
$0<\epsilon\ll 1$. They can be written in the form of a conservation law
\eqa\label{tr2}
&&
u_t+a(u) u_x +\epsilon^2 \pal_x \left\{  b_1(u) u_{xx} + b_2(u) u_x^2+\epsilon \left[ b_3(u) u_{xxx} + b_4(u) u_{xx} u_x +b_5(u) u_x^3\right]
\right.
\nn\\
&&\left.
+\epsilon^2 \left[ b_6(u) u_{xxxx} + b_7(u) u_{xxx} u_x + b_8(u) u_{xx}^2 +b_9(u) u_{xx} u_x^2 + b_{10}(u) u_x^4\right]\right\}=0,
\eeqa
where the coefficients $b_1(u)$, \dots, $b_{10}(u)$ are smooth 
functions satisfying certain constraints following from the existence of a Hamiltonian representation
$$
u_t+\pal_x \frac{\delta H}{\delta u(x)} =0, 
$$
(see Corollary \ref{cor22} below). Here and below we use the notation
$$
\pal_x =\frac{\pal}{\pal x}.
$$
This class of equations contains important equations 
as the Korteweg-de Vries (KdV) equation $u_t+6uu_x+\epsilon^2u_{xxx}=0$  and its  generalizations, the Kawahara 
equation and the Camassa-Holm equation in an asymptotic sense (see  \cite{dub06}). 

Up to certain equivalencies the Hamiltonian regularizations of (\ref{tr1}) 
are characterized by two free functions $c(u)$ and $p(u)$:
\beq\label{cau}
u_t+a(u) u_x +{\epsilon^2} \pal_x \left[  c\, a' u_{xx} +\frac12(c\, a')' u_x^2\right] +\epsilon^4 \pal_x \left[ \left( 2 p\, a' +\frac35{c^2 a''}\right) u_{xxxx}+\dots\right]=0.
\eeq
Two equations of the form \eqref{cau} with the same invariants $c(u)$ and $p(u)$ commute, up to order ${\mathcal O}(\epsilon^6)$
$$
\left( u_t\right)_{\tilde t} -\left( u_{\tilde t}\right)_t={\mathcal O}\left(\epsilon^6\right)
$$
where, for an arbitrary function $\tilde a=\tilde a(u)$
$$
u_{\tilde t}+\tilde a(u) u_x +{\epsilon^2} \pal_x \left[  c\, \tilde a' u_{xx} +\frac12(c\, \tilde a')' u_x^2\right] +\epsilon^4 \pal_x \left[ \left( 2 p\, \tilde a' +\frac35{c^2 \tilde a''}\right) u_{xxxx}+\dots\right]=0.
$$

In this paper 
the analysis of  \cite{dub06} up to order $\epsilon^{4}$ is extended 
to higher orders of $\epsilon$. Our analysis suggests that the only obstruction 
to the functions $c(u)$ and $p(u)$ by the condition of integrability 
is the condition that $c(u)$ must not vanish. 


We then proceed to the study of the \emph{critical behaviour} of solutions to \eqref{tr2}. Namely,
let $(x_c, t_c, u_c)$ be a point of \emph{gradient catastrophe} of a 
solution $u^0(x,t)$ to \eqref{tr1} specified by an initial value $u^0(x,0)=\phi(x)$. This means that the solution is a 
smooth function of $(x,t)$ for sufficiently small $|x-x_c|$ and 
$t-t_c<0$. Moreover there exists the limit
$$
\lim_{x\to x_c, \, t\to t_c-0} u^0(x,t)=u_c,
$$
but the derivatives $u_x^0(x,t)$, $u_t^0(x,t)$ blow up at the point.
The \emph{Universality Conjecture} of \cite{dub06} says that, up to shifts, Galilean transformations and 
rescalings, the behavior at the point of gradient catastrophe of a solution to \eqref{tr2} with the \emph{same} $\epsilon$-independent initial data $\phi(x)$ essentially depends neither on the choice 
of the generic solution nor on the choice of the generic equation.
Moreover, the generic solution near this point $(x_c, t_c, u_c)$ is given by
\begin{equation}
\label{univer}
u(x,t,\epsilon)\simeq u_c +\alpha\,\epsilon^{2/7} U \left(\dfrac{
 x- x_c-a_0 (t-t_c)}{\beta\,\epsilon^{6/7}};
\dfrac{t-t_c}{\gamma\,\epsilon^{4/7}}\right) +O\left( \epsilon^{4/7}\right),
\end{equation}
where  $a_0=a(u_c)$ and the constants $\alpha$, $\beta$, $\gamma$ depend on the choice
of the generic equation and the solution
\eqa\label{abc}
&&
\alpha=\left(\frac{12 b_1^0}{a_0' k^2}\right)^{1/7},
\nn\\
&&
\beta=\left( \frac{12^3 k\, (b_1^0)^3}{{a_0'}^3}\right)^{1/7},
\\
&&
\gamma=\left( \frac{12^2 k^3 (b_1^0)^2}{{a_0'}^9}\right)^{1/7}.
\nn
\eeqa
Here $a_0'=a'(u_c)$, $b_1^0=b_1(u_c)$; it is assumed that $b_1^0\neq 0$. The constant $k$ in these formula is inverse proportional to the ``strength" of the breakup of the dispersionless solution $u^0(x,t)$
\beq\label{ko}
k=-6 \lim_{x\to x_c} \frac{x-x_c}{\left( u^0(x, t_c) -u_c\right)^3}
\eeq
where we assume that $k\neq 0$ (another genericity hypothesis) and $a_0' k>0$.

 The function
$U=U(X; T)$, $(X,T) \in\mathbb R^2$, is defined as the unique real smooth  
solution to the fourth order ODE  \cite{bmp}, \cite{kapaev}
\begin{equation}\label{PI2}
X=T\, U -\left[ \frac16U^3 + \frac1{24} U_{X}^2 +  \frac1{12}U\, U_{XX}  +\frac1{240} U_{XXXX}\right],
\end{equation}
which is  the second member 
of the Painlev\'e I hierarchy. We will call this equation PI2. 

The relevant solution  is characterized by the asymptotic behavior
\begin{equation}
\label{PI2asym}
	U(X,T)=-(6X)^{\frac{1}{3}}- 
	\dfrac{2^{2/3}T}{(3X)^{\frac13}}+O(X^{-\frac53}),\quad X\to \pm \infty,
\end{equation}
for each fixed $T\in\mathbb{R}$.
The existence of a smooth solution of (\ref{PI2}) for all  real $X,T$ 
satisfying (\ref{PI2asym}) has been proved by Claeys and Vanlessen \cite{cl1}.

Observe that the principal term of the asymptotics \eqref{univer} depends only on the order $\epsilon^2$ regularization. In the present paper we will numerically analyze, in particular, the influence of the higher order corrections\footnote{One should also take into account \cite{dub06} that the actual small parameter of the expansion \eqref{univer} is 
$$
\left( 12 b_1^0 \epsilon^2\right)^{1/7}.
$$
In other words the asymptotic expansion \eqref{univer} makes sense 
only under the assumption
$$
b_1^0 \epsilon^2 \ll \frac1{12}
$$
in agreement with \cite{bk}.} on the local behaviour of solutions to \eqref{tr2} near the point of catastrophe.

First numerical tests of the PI2 asymptotic description of the critical point 
for the class of PDE in \cite{dub06} have been presented in 
\cite{gk08} for the KdV and the Camassa-Holm equation.
In \cite{cg} a rigourous proof of the asymptotic behaviour (\ref{univer}) has been obtained for the KdV equation.

In this paper we generalize the numerical investigation of \cite{gk08} to a larger class of equations 
  which include the  generalized KdV equation, the  Kawahara equations with a dispersion of fifth order 
and the second equation in the KdV hierarchy. 

We comment on the formation of blow up  and  on the 
role of integrability in  the formation of oscillatory regions. In particular we show the differences in the formation of dispersive shock waves between integrable and non-integrable cases.
The KdV equation has been extensively studied numerically in \cite{gk06}.

Then we show numerically  that the solution of the dispersive equation (\ref{tr2}) converges to the solution of the dispersionless equation (\ref{tr1}) away from the point of gradient catastrophe at a rate of order $\epsilon^2$. Finally we show that the solution of the dispersive equation (\ref{tr2})  is well approximated as a series in even power of $\epsilon$ in terms of  the solution of the dispersionless equation  (\ref{tr1}) up to order $\epsilon^4$ by the so called quasi-triviality transformation \cite{dub06} away from the point of gradient catastrophe.
Such an approximation has already  been obtained for conservation 
laws with positive viscosity \cite{GX92}.  Furthermore the existence 
of an expansion in even powers of $\epsilon$ has already appeared and 
been proved in  the context of large $N$ expansions in Hermitian  matrix models \cite{BI},\cite{EM}.

The paper is organized as follows. In sect.~2 we briefly review the 
results of \cite{dub06}. In sect.~3 we discuss higher order in 
$\epsilon$ regularizations of (\ref{tr1}) and obstructions on the 
function $c(u)$ by the condition of integrability.  A numerical 
study of the applicability of the conjecture to generalized KdV 
equations is given in sect.~4. We also comment on the possibility of 
blowup. In sect.~5 the conjecture is tested numerically for equations 
with high order dispersion as the Kawahara equation. Differences in the formation of rapid 
oscillations in the solutions   to 
integrable and non-integrable equations are studied.
Details about the used numerical methods are given in the appendix.

\section{Hamiltonian PDEs and their invariants}

In  this paper we mainly study scalar Hamiltonian PDEs of the order at most five. They are written in the form of a conservation law
\beq\label{cl1}
u_t +\pal_x\varphi (u, \epsilon\, u_x, \epsilon^2u_{xx}, 
\epsilon^3u_{xxx}, \epsilon^4u_{xxxx})=0,
\eeq
where
\eqa\label{cl2}
&&
\varphi =\frac{\delta H}{\delta u(x)},
\\
&&
H=\int h(u, \epsilon\,u_x, \epsilon^2u_{xx})\, dx.
\nn
\eeqa
Recall that the Euler--Lagrange derivative is defined by
\beq\label{el}
\frac{\delta H}{\delta u(x)}=\frac{\pal h}{\pal u} -\pal_x \frac{\pal h}{\pal u_x}+\pal_x^2 \frac{\pal h}{\pal u_{xx}}-\dots.
\eeq
Here and in the sequel the integral of a differential polynomial is understood, in the spirit of \emph{formal calculus of variations}, as the equivalence class of the polynomial modulo the image of the operator of total $x$-derivative
\beq\label{delx}
\pal_x h =u_x \frac{\pal h}{\pal u}+u_{xx} \frac{\pal h}{\pal u_x}+\dots.
\eeq
It is worthwhile to recall that a differential polynomial $p(u; u_x, \dots, u^{(m)})$ belongs to ${\rm Im}\, \pal_x$ \emph{iff}
\beq\label{crit}
\frac{\delta P}{\delta u(x)}=0, \quad P=\int p(u; u_x, \dots, u^{(m)})\, dx.
\eeq
The Poisson bracket of two local functionals $H$, $F$ associated with \eqref{cl1}, \eqref{cl2},  is a local functional of the form
\beq\label{gfz1}
\{ H, F\} =\int \frac{\delta H}{\delta u(x)} \frac{d}{dx} \frac{\delta F}{\delta u(x)}\, dx
\eeq

\begin{lemma} Equation \eqref{cl1} can be written in the Hamiltonian form \eqref{cl2} \emph{iff} the function $\varphi$ satisfies the following two constraints
\eqa\label{helm1}
&&
\frac{\pal \varphi}{\pal u_x} = \pal_x \left[ \frac{\pal \varphi}{\pal u_{xx}}-\frac12 \pal_x\frac{\pal \varphi}{\pal u_{xxx}}\right]
\nn\\
&&
\\
&&
\frac{\pal \varphi}{\pal u_{xxx}} =2 \pal_x \frac{\pal \varphi}{\pal u_{xxxx}}.
\nn
\eeqa
\end{lemma}

\pf According to the classical Helmholtz criterion (see in \cite{dt}) the function $\varphi(u, u_x, u_{xx}, \dots)$ can be locally represented as the variational derivative of some functional $H=\int h(u, u_x, \dots)\, dx$ \emph{iff} it satisfies the following system of constraints
\beq\label{helm0}
\frac{\pal \varphi}{\pal u^{(i)}} =(-1)^i \sum_{m\geq 0} \frac{(m+i)!}{i! ~m!} (-\pal_x)^m \frac{\pal \varphi}{\pal u^{(i+m)}}, \quad i=0, \, 1, \, \dots.
\eeq
For the particular case under consideration the equations \eqref{helm0} reduce to \eqref{helm1}. \epf

Applying the Lemma to a PDE \eqref{cl1} written in the form of the weak dispersion expansion one arrives at

\begin{cor}\label{cor22} The equation
\eqa\label{ur0}
&&
u_t+a(u) u_x +\pal_x \left\{\epsilon \, b_0(u) u_{x} +\epsilon^2 \left[b_1(u) u_{xx} + b_2(u) u_x^2\right]+\epsilon^3 \left[ b_3(u) u_{xxx} + b_4(u) u_{xx} u_x +b_5(u) u_x^3\right]
\right.
\nn\\
&&\left.
+\epsilon^4 \left[ b_6(u) u_{xxxx} + b_7(u) u_{xxx} u_x + b_8(u) u_{xx}^2 +b_9(u) u_{xx} u_x^2 + b_{10}(u) u_x^4\right]\right\}=0
\eeqa
is Hamiltonian \emph{iff} the coefficients $b_0$, \dots, $b_{10}$ satisfy
\eqa
&&
b_0=0,
\nn\\
&&
b_2=\frac12 b_1',
\nn\\
&&
b_3=0,
\nn\\
&&
b_5=\frac13 b_4',
\nn\\
&&
b_7=2 b_6',
\nn\\
&&
b_8=\frac32 b_6',
\nn\\
&&
b_{10}=\frac14  b_9'.
\nn
\eeqa
\end{cor}

The Hamiltonian equations \eqref{cl1} are considered modulo canonical transformations written in the form of a time-$\epsilon$ shift 
\beq\label{can1}
u(x) \mapsto \tilde u(x)=u(x) +\epsilon\{ u(x), K\} +\frac{\epsilon^2}{2!} \left\{\{ u(x), K\}, K\right\} +\dots
\eeq
generated by a Hamiltonian
\beq\label{can2}
K=\int k(u, \epsilon\, u_x, \dots)\, dx.
\eeq
The transformations \eqref{can1} preserve the canonical form of the Poisson bracket \eqref{gfz1}.
Two Hamiltonian equations are called \emph{equivalent} if they are related by a canonical transformation of the form \eqref{can1}, \eqref{can2}.
For example, the degree 3 terms in a Hamiltonian PDE of the form \eqref{ur0} can be eliminated by a transformation \eqref{can1} if $a'(u) \neq 0$.  Indeed, it suffices to choose the generating Hamiltonian in the form
$$
K=\int \epsilon^2 \frac{b_4(u)}{6 a'(u)} u_x^2\, dx.
$$

The following Lemma describes a normal form of Hamiltonians of order 4 (cf. \cite{dub06}) with respect to transformations \eqref{can1}.

\begin{lemma} Any Hamiltonian equation of the form \eqref{ur0} with $a'(u)\neq 0$ is equivalent to 
\eqa\label{urc}
&&
u_t +a(u) u_x +\epsilon^2 \pal_x \left[ b_1 u_{xx} +\frac12 b_1' u_x^2 +\epsilon^2 \left( b_6 u_{xxxx} + 2 b_6' u_{xxx} u_x +\frac32 b_6' u_{xx}^2 
\right.\right.
\nn\\
&&
\\
&&\left.\left.
\quad\quad\quad\quad\quad\quad\quad\quad +b_9 u_{xx} u_x^2 +\frac14 b_9'  u_x^4\right)\right]=0.
\nn
\eeqa
The Hamiltonian PDEs \eqref{urc} and
\eqa\label{urc1}
&&
u_t +\tilde a(u) u_x +\epsilon^2 \pal_x \left[ \tilde b_1 u_{xx} +\frac12 \tilde b_1' u_x^2 +\epsilon^2 \left( \tilde b_6 u_{xxxx} + 2 \tilde b_6' u_{xxx} u_x +\frac32 \tilde b_6' u_{xx}^2 
\right.\right.
\nn\\
&&
\\
&&\left.\left.
\quad\quad\quad\quad\quad\quad\quad\quad +\tilde b_9 u_{xx} u_x^2 +\frac14  b_9'  u_x^4\right)\right]=0.
\nn
\eeqa
are equivalent \emph{iff}
\beq
\tilde a=a, \quad \tilde b_1=b_1, \quad \tilde b_6=b_6.
\eeq
\end{lemma}

\pf We have already proved that the coefficients of degree 3 in $\epsilon$ can be eliminated by a canonical transformation of the form \eqref{can1}. One can easily see that the coefficients $a$, $b_1$ and $b_6$ are invariant with respect to these transformations. Two Hamiltonians of the form
$$
H=\int \left[ f-\frac{\epsilon^2}2 b_1 u_x^2 +\frac{\epsilon^4}2 b_6 u_{xx}^2 -\frac{\epsilon^4}{12} b_9 u_x^4\right]\, dx
$$
and
$$
\tilde H=\int \left[ f-\frac{\epsilon^2}2 b_1 u_x^2 +\frac{\epsilon^4}2 b_6 u_{xx}^2 -\frac{\epsilon^4}{12} \tilde b_9 u_x^4\right]\, dx
$$
generating the flows \eqref{urc} and \eqref{urc1} with the same coefficients $\tilde a=a$, $\tilde b_1=b_1$, $\tilde b_6=b_6$ but with different $\tilde b_9\neq b_9$ are related by a canonical transformation \eqref{can1} with
$$
K=\frac{\epsilon^3}{24} \int \frac{\tilde b_9-b_9}{a'}\, u_x^3 dx.
$$ \epf

Thus the coefficients $a$, $b_1$, $b_6$ are \emph{invariants} of the Hamiltonian PDE \eqref{urc}.

As it was discovered in \cite{dub06}, any Hamiltonian PDE of the form \eqref{urc} is integrable at the order $\epsilon^4$ approximation. More precisely, assuming $a'\neq 0$ let us replace the invariants $b_1=b_1(u)$ and $b_6=b_6(u)$ with
\beq\label{inva}
c=\frac{b_1}{a'}, \quad p=\frac{b_6}{2 a'} -\frac3{10} b_1^2 \frac{a''}{{a'}^3}.
\eeq
Then the equation \eqref{urc} is equivalent to the PDE
\beq\label{ur}
u_t+a(u) u_x +\epsilon^2 \pal_x \left[  c\, a' u_{xx} +\frac12(c\, a')' u_x^2\right] +\epsilon^4 \pal_x \left[ \left( 2 p\, a' +\frac35 c^2 a''\right) u_{xxxx}+\dots\right]=0
\eeq 
with the Hamiltonian
\eqa\label{cub2}
&&
H_f=\int\left\{ f -\frac{\epsilon^2}{2} c\, f''' u_x^2 +\epsilon^4 \left[ \left( p \, f''' +\frac3{10} c^2 f^{(4)}\right) u_{xx}^2
\right.\right.
\nn\\
&&
\\
&&\left.\left.
-\frac16\left( \frac{ 3 c\, c'' f^{(4)} +3c\, c' f^{(5)}+c^2 f^{(6)}}{4}+p' f^{(4)} +p\, f^{(5)} 
   \right) u_x^4\right]\right\}\, dx
\nn
\eeqa
where, as above,
$$
f''(u) = a(u).
$$
The approximate integrability means that, fixing the functional 
parameters $c=c(u)$, $p=p(u)$ one obtains a family of Hamiltonians satisfying
\beq\label{cub3}
\{ H_f, H_g\}={\mathcal O}\left( \epsilon^6\right)
\eeq
for an arbitrary pair of smooth functions $f=f(u)$, $g=g(u)$. 
In particular choosing $f(u)=\frac16 u^3$ one obtains the Hamiltonian
\beq\label{hopf1}
H=\int\left[ \frac{u^3}6 - \epsilon^2 \frac{c(u)}{2} u_x^2
+\epsilon^4  p(u) u_{xx}^2\right]\, dx
\eeq
of a general order 4 dispersive regularization of the Hopf equation
\beq\label{hopf2}
u_t+u\, u_x +\epsilon^2 \pal_x \left[ c\, u_{xx} +\frac12 c' u_x^2\right] +\epsilon^4 \pal_x \left[ 2 p\, u_{xxxx} +4 p' u_{xxx} u_x +3 p' u_{xx}^2 + 2 p'' u_{xx} u_x^2\right] =0
\eeq
introduced in \cite{dub06}\footnote{In the present paper we use a different normalization $c(u) \mapsto 12 c(u)$.}.

More generally, we call a perturbation 
$$
H=H_0 +\epsilon\, H_1 +\epsilon^2 H_2+\dots
$$
of the Hopf Hamiltonian
$$
H_0=\int \frac{u^3}6\, dx
$$
\emph{$N$-integrable} if, for any smooth function $f=f(u)$ there exists a perturbed Hamiltonian
$$
H_f=H_f^0+\sum_{k\geq 1} \epsilon^k H_f^k
$$
such that for $f=\frac{u^3}6$ the Hamiltonian $H_f$ coincides with $H$ and, moreover, for any pair of functions $f$, $g$ the Hamiltonians $H_f$, $H_g$ satisfy
$$
\{ H_f, H_g\}={\mathcal O}\left(\epsilon^{N+1}\right).
$$
For example, the perturbed Hamiltonian \eqref{hopf1} is 5-integrable. The commuting Hamiltonians have the form \eqref{cub2}. In the next section we will discuss the problem of constructing higher integrable perturbations of \eqref{hopf1}.

\section{On obstacles to integrability}\label{sec3}

We will now study the possibility to extend the commuting Hamiltonians \eqref{cub2} to the next order of the perturbative expansion.

\begin{theorem}\label{the01} 1) Any order 6 perturbation of the cubic Hamiltonian $H_0=\int \frac{u^3}6\, dx$ can be represented in the form
\beq\label{pert6}
H=\int\left[ \frac{u^3}6 - \frac{\epsilon^2}2 c(u) u_x^2
+\epsilon^4  p(u) u_{xx}^2-\epsilon^6\left(\alpha(u) u_{xxx}^2+\beta(u) u_{xx}^3\right) \right]\, dx
\eeq
Such a perturbation is 7-integrable for arbitrary functional parameters $c=c(u)$, $p=p(u)$, $\alpha=\alpha(u)$, $\beta=\beta(u)$.

\noindent 2) The perturbation \eqref{pert6} can be extended to a 9-integrable one \emph{iff} $c(u)\neq 0$ and
\beq\label{alpha1}
 \alpha=\frac1{28}\left[80\frac{p^2}{c} - 67\,p\,c' + 33\,c\,p'+
 12\,c\,{c'}^2  - 
 9\,c^2\,c''\right].
\eeq

\end{theorem}

\pf 
A general order 6 perturbation of the cubic Hamiltonian $H_0$ must have the form
$$
H=\int\left\{\frac{u^3}6 -\frac{\epsilon^2}2 {c(u)} u_x^2
+\epsilon^4  p(u) u_{xx}^2-\epsilon^6\left[\alpha(u) u_{xxx}^2 +\beta(u) u_{xx}^3 +\gamma(u) u_{xx}^2 u_x^2 + \delta(u) u_x^6\right]\right\}\, dx.
$$
The last two terms can be eliminated by a canonical transformation
$$
H\mapsto H-\epsilon\{ H, F\} +\dots
$$
with
$$
F=\int \epsilon^5 \left( \frac16 \gamma(u) u_{xx}^2 u_x +\frac14 \delta(u)u_x^5\right)\, dx.
$$
For an arbitrary function $f=f(u)$ the density of a Hamiltonian
$$
H_f =\int h_f\, dx
$$
commuting with \eqref{pert6} modulo ${\mathcal O}\left( \epsilon^6\right)$ must have the form
\eqa\label{ham6}
&&
h_f=f -\frac{\epsilon^2}{2} c\, f''' u_x^2 +\epsilon^4 \left[ \left( p \, f''' +\frac3{10} c^2 f^{(4)}\right) u_{xx}^2
-\frac16\left( p' f^{(4)} +\frac34 c\, c'' f^{(4)} +p\, f^{(5)} 
\right.\right.
\nn\\
&&
\\
&&\left.\left.
+\frac34 c\, c' f^{(5)} +\frac14 c^2 f^{(6)}\right) u_x^4\right]
-\epsilon^6 \left[ \alpha_f(u) u_{xxx}^2 + \beta_f(u) u_{xx}^3 +\gamma_f(u) u_{xx}^2 u_x^2 +\delta_f(u) u_x^6\right]
\nn
\eeqa
with some smooth functions $\alpha_f=\alpha_f(u)$, $\beta_f=\beta_f(u)$, $\gamma_f=\gamma_f(u)$, $\delta_f=\delta_f(u)$  depending on $f$. From the commutativity
$$
\{ H, H_f\} ={\mathcal O}\left( \epsilon^7\right)
$$
one uniquely determines these coefficients
$$
\alpha_f= \alpha\, f''' +\left( \frac87 c\, p +\frac3{70}{c^2 c' }\right) f^{(4)} +\frac9{70}{c^3 f^{(5)}}
$$
\eqa
&&
\beta_f=\beta\, f'''-\left( \frac32 \alpha+\frac{253 p\, c'+169 c\, p' }{168}+\frac{c\, {c'}^2 }{35}   +\frac5{56} c^2 c'' \right) f^{(4)}
-\left( \frac{29}{21} c\, p+\frac{31}{70} c^2 c' \right)f^{(5)} -\frac{c^3 f^{(6)}}{7}
\nn
\eeqa
\eqa
&&
\gamma_f=\left(\frac37\beta -\frac67 \alpha'  +\frac3{35}({c'}^3-c^2 c''' -3c\, c' c'') + c' p'  -\frac{47}{14} p\, c''   - c\, p''  \right)f^{(4)}
\nn\\
&&
-\left(2 \alpha+\frac{37}{14} p\, c' +\frac3{35}({c\, {c'}^2+11 c^2 c'' }) +\frac87 c\, p'  \right)f^{(5)} -\frac1{14}\left({23} c\, p+9{c^2 c'}\right) f^{(6)}  -\frac3{20}{c^3 f^{(7)}}
\nn
\eeqa
\eqa
&&
\delta_f=\left(\frac1{10} p' c''' +\frac{10 c\, c'' c'''+7 c\, c' c^{(4)}+c^2 c^{(5)}}{40} +\frac2{15} p\, c^{(4)}  +\frac1{60} c\, p^{(4)} \right) f^{(4)}
\nn\\
&&
+\left(\frac1{15} \alpha'' +\frac1{5} p' c'' +\frac3{40}{c\, {c''}^2} +\frac3{10} p\, c''' +\frac{c\, c' c'''}{10} +\frac1{15} c\, p''' +\frac{c^2 c^{(4)}}{15}\right) f^{(5)}
\nn\\
&&
+\left( \frac2{15} \alpha' + \frac2{15} c' p' +\frac1{3} p\, c'' +\frac{7c\, c' c''+3c^2 c'''} {40}  +\frac1{10} c\, p'' \right) f^{(6)}
\nn\\
&&
+\left( \frac1{15} \alpha +\frac1{6} p\, c' +\frac{c\, {c'}^2}{16}+\frac1{10} c\, p' +\frac3{40}{c^2 c''}\right) f^{(7)}
+\left(\frac1{20}c\, p +\frac3{80}{c^2 c'}\right) f^{(8)}
+\frac{c^3}{240} f^{(9)}
\nn
\eeqa
Thus the resulting Hamiltonian $H_f$ satisfies
$$
\{ H, H_f\} ={\mathcal O}\left( \epsilon^8\right).
$$
It is not difficult to also verify the commutativity
$$
\{ H_f, H_g\} ={\mathcal O}\left( \epsilon^8\right)
$$
for an arbitrary pair of functions $f=f(u)$, $g=g(u)$.

Let us now analyze the possibility of extension to a commutative family of order 8. We add to \eqref{pert6} terms of the form
\eqa
&&
H\mapsto \tilde H=H+\int \epsilon^8 \left[ A_1 u_x^8 + A_2 u_x^4 u_{xx}^2 +A_3 u_x^2 u_{xx}^3 +A_4 u_{xx}^4
\right.
\nn\\
&&\left.
 +A_5 u_x^2 u_{xxx}^2 +A_6 u_{xx} u_{xxx}^2 +A_7  u_{xxxx}^2\right]\, dx
 \nn
\eeqa
and to \eqref{ham6} a similar expression
\eqa
&&
H_f\mapsto \tilde H_f=H_f+\int\epsilon^8 \left[ B_1 u_x^8 + B_2 u_x^4 u_{xx}^2 +B_3 u_x^2 u_{xx}^3 +B_4 u_{xx}^4 
\right.
\nn\\
&&
\left.
+B_5 u_x^2 u_{xxx}^2 +B_6 u_{xx} u_{xxx}^2 +B_7  u_{xxxx}^2\right]\, dx.
\nn
\eeqa
Here $A_1$, \dots, $A_7$, $B_1$, \dots, $B_7$ are some functions of $u$.
The goal is to meet the condition
\beq\label{int9}
\{ \tilde H, \tilde H_f\} ={\mathcal O}\left( \epsilon^9\right).
\eeq
The order 8 terms in the bracket \eqref{int9} are represented by a differential polynomial of degree 9.
From the vanishing of the coefficient of $u^{(8)} u_x$ it follows that
\eqa
&&
B_7=A_7\,f''' + \left( \frac{10}9\,\alpha\,c + 
 \frac{10}9 p^2 + 
 \frac{10}{63}\,c\,c' p- 
 \frac1{210}c^2\,{c'}^2 + 
 \frac1{21}c^2\,p' + 
 \frac1{70}c^3\,c''\right)\,f^{(4)}
 \nn\\
 && + 
\left( \frac57\,c^2\,p + 
 \frac3{70}c^3\,c'\right)\,f^{(5)} + 
 \frac3{70}c^4\,f^{(6)}.
 \nn
\eeqa 
Next, from the vanishing of the coefficient of $u^{(6)} u_{xxx}$ we get \eqref{alpha1}. 
 
Further calculations allow one to determine $B_6$ from the coefficient of $u^{(6)} u_{xx} u_x$,
 $B_5$ from the coefficient of $u^{(6)} u_x^3$, $B_4$ from the coefficient of $u^{(4)} u_{xx}^2 u_x$,
$B_3$ from the coefficient of
$u^{(4)} u_{xx} u_x^3$,
$B_2$ from the coefficient of $u^{(4)} u_x^5$,
and, finally, $B_1$ from the coefficient of $u_{xx} u_x^7$. All these coefficients are represented by linear differential operators of order at most 12 acting on the arbitrary function $f=f(u)$. The coefficients of these operators depend linearly on $A_1$, \dots, $A_7$ and their $u$-derivatives and also on $c(u)$ and $p(u)$ and their derivatives. The explicit formulae are rather long; they will not be given here. As above one can verify validity of the identity
$$
\left\{ \tilde H_f, \tilde H_g\right\} ={\mathcal O}\left( \epsilon^{10}\right)
$$
for any pair of functions $f(u)$, $g(u)$. \epf


\begin{cor} Let $p(u)$ be an arbitrary non-vanishing function. Then the Hamiltonian
\beq\label{obst}
H=\int\left[ \frac{u^3}6 
+\epsilon^4  p(u) u_{xx}^2\right]\, dx
\eeq
cannot be included into a 9-integrable family.
\end{cor}



\section{Quastriviality transformations and perturbative solutions}

In this section we will develop  a perturbative technique for constructing monotone solutions to the equations of the form \eqref{hopf2} for sufficiently small time $t$. This technique is based on the so-called \emph{quasitriviality transformation} \cite{dub06} expressing solutions to the perturbed equation \eqref{hopf2} in terms of solutions to the unperturbed equation. 

To explain the basic idea let us consider the equation
\eqa\label{eq1}
&&
u_t+u\, u_x + \epsilon^2\pal_x \left( c\, u_{xx} +\frac12 c' u_x^2\right)+\dots=u_t+\pal_x \frac{\delta H}{\delta u(x)}=0
\\
&&
H=\int \left[ \frac12 u^3 -\frac{\epsilon^2}2 c\, u_x^2+\dots\right]\, dx.
\nn
\eeqa
The quasitriviality transformation for this equation
\beq\label{quasi1}
v\to u=v +\epsilon^2 \left[ \frac{c}2\,\left( \frac{v_{xxx}}{ v_x}-\frac{ v_{xx}^2}{ v_x^2}\right) +c' v_{xx} +\frac12 c'' v_x^2\right]+{\mathcal O}(\epsilon^4)
\eeq
 is generated by the Hamiltonian
 \beq\label{quasi2}
 K=-\frac{\epsilon}2 \int c\, v_x \log v_x \, dx +{\mathcal O}(\epsilon^3),
 \eeq
 $$
 u=v+\epsilon\{ v(x), K\} +\frac{\epsilon^2}{2!} \{\{ v(x), K\}, K\}+\dots.
 $$
Substituting into eq. \eqref{eq1} one obtains a function $u(x,t;\epsilon)$ satisfying \eqref{eq1} up to terms of order $\epsilon^4$. Indeed, one can easily derive the following expression for the discrepancy
\eqa
&&
\epsilon^{-4}\left[u\, u_x + \epsilon^2\pal_x \left( c\, u_{xx} +\frac12 c' u_x^2\right)-u_t\right]=
\nn\\
&&
=c^2 \left( \frac{23 {v_{xx}}^5}{2 {v_{x}}^5}-\frac{115 {v_{xx}}^3 {v_{xxx}}}{4 \
{v_{x}}^4}+\frac{39 {v_{xx}}^2 {v_{xxxx}}}{4 {v_{x}}^3}+\frac{57 \
{v_{xx}} {v_{xxx}}^2}{4 {v_{x}}^3}-\frac{5 {v_{xx}} {v_{xxxxx}}}{2 \
{v_{x}}^2}-\frac{19 {v_{xxx}} {v_{xxxx}}}{4 {v_{x}}^2}+\frac{{v_{xxxxxx}}}{2 \
{v_{x}}}
\right)
\nn\\
&&
+c\, c' \left( -\frac{35 {v_{xx}}^4}{4 {v_{x}}^3}+\frac{19 {v_{xx}}^2 \
{v_{xxx}}}{{v_{x}}^2}-\frac{7 {v_{xx}} {v_{xxxx}}}{{v_{x}}}-\frac{23 \
{v_{xxx}}^2}{4 {v_{x}}}+\frac{7 {v_{xxxxx}}}{2}
\right)
\nn\\
&&
+c\, c'' \left(\frac{3 {v_{xx}}^3}{2 {v_{x}}}+\frac{13 {v_{x}} {v_{xxxx}}}{2}+3 \
{v_{xx}} {v_{xxx}}
\right)
+c\, c''' \left(\frac{15 {v_{x}}^2 {v_{xxx}}}{2}+8 {v_{x}} {v_{xx}}^2
\right)+\frac{11}2 c\, c^{(4)} v_x^3 v_{xx}+\frac12 c\, c^{(5)} v_x^5
\nn\\
&&
+{c'}^2 \left(\frac{3 {v_{xx}}^3}{2 {v_{x}}}+4 {v_{x}} {v_{xxxx}}+\frac{{v_{xx}} \
{v_{xxx}}}{2}
\right)+ c' c'' \left(\frac{21 {v_{x}}^2 {v_{xxx}}}{2}+10 {v_{x}} {v_{xx}}^2\right)+9 c' c''' v_x^3 v_{xx}+c' c^{(4)} v_x^5
\nn\\
&&
+5 {c''}^2 v_x^3 v_{xx}+\frac54 c'' c''' v_x^5+{\mathcal O}(\epsilon^2)
\nn
\eeqa
Note that the \emph{same} quasitriviality transformation works for solutions $v=v(x,t)$ to the nonlinear transport equation
$$
v_t+a(v) v_x=0
$$
transforming it to solutions, modulo ${\mathcal O}(\epsilon^4)$, to the perturbed equation \eqref{tr1}.

Denote by $\phi(x)=v(x,0)$ the initial data for the Hopf equation. The initial value of solution $u(x,t;\epsilon)$ given by the formula  \eqref{quasi1} differs from $\phi(x)$:
\beq\label{init1}
u(x,0;\epsilon) =\phi +\epsilon^2 \left[ \frac{c}2\,\left( \frac{\phi_{xxx}}{ \phi_x}-\frac{ \phi_{xx}^2}{ \phi_x^2}\right) +c' \phi_{xx} +\frac12 c'' \phi_x^2\right]+{\mathcal O}(\epsilon^4).
\eeq
In order to solve the Cauchy problem for \eqref{eq1} with the \emph{same} initial data $u(x,0;\epsilon)=\phi(x)$ one can use the following trick. Let us consider the solution $\tilde v =\tilde v(x,t;\epsilon)$ to the Hopf equation with the $\epsilon$-dependent initial data
\beq\label{init2}
\tilde v(x,0;\epsilon) =\phi -\epsilon^2 \left[ \frac{c}2\,\left( \frac{\phi_{xxx}}{ \phi_x}-\frac{ \phi_{xx}^2}{ \phi_x^2}\right) +c' \phi_{xx} +\frac12 c'' \phi_x^2\right].
\eeq
Such a solution can be represented in the form
\beq\label{init3}
\tilde v(x,t;\epsilon) =v(x,t)+\epsilon^2 w(x,t)+{\mathcal O}(\epsilon^4)
\eeq
where the function $w(x,t)$ has to be determined from the equation
\beq\label{init4}
\Phi'(w) -w\, t = \left[\frac{c(v)}2 \frac{2 {\Phi''}^2(v) -\Phi'(v) \Phi'''(v)}{{\Phi'}^3(v)} -c'(v)\frac{\Phi''(v)}{{\Phi'}^2(v)}+\frac{c''(v)}{2 \Phi'(v)}\right]_{v=v(x,t)}.
\eeq
Here $\Phi(v)$ is the function inverse to $\phi(x)$.
Applying the quasitriviality transformation to the solution $\tilde v(x,t; \epsilon)$ one obtains a function
\beq\label{quasidue}
u(x,t;\epsilon) =v+\epsilon^2 w +\epsilon^2 \left[ \frac{c}2\,\left( \frac{v_{xxx}}{ v_x}-\frac{ v_{xx}^2}{ v_x^2}\right) +c' v_{xx} +\frac12 c'' v_x^2\right]
\eeq
satisfying equation \eqref{eq1} modulo terms of order $\epsilon^4$ with the initial data
\beq\label{init5}
u(x,0;\epsilon)=\phi(x) +{\mathcal O}(\epsilon^4).
\eeq

\section{Generalized KdV equations}
In this section we will first study the role of the function $a(u)$ 
in (\ref{tr1}) on the validity of the conjecture. This is done for the
generalized KdV equations having the form
\begin{equation}
    u_{t}+a(u)u_{x}+\epsilon^{2}u_{xxx}=0.
    \label{genKdV}
\end{equation}
We will assume that $a(u)$ is 
monotonic in an open neighborhood of each critical point.  The functional parameters $c(u)$ and $p(u)$ in (\ref{cub2})  are given by
\beq\label{kdv-cp}
c(u)=\frac{1}{a'(u)}, \quad p(u)=-\frac3{10} \frac{a''(u)}{{a'(u)}^3}.
\eeq

The basic idea of the PI2 approach to the breakup behavior is that 
the equation behaves in this case approximately as the KdV equation. 
We will test this assumption first for  $a(u)$ of the form 
$a(u)=6u^{n}$, $n\in \mathbb{N}$.

\subsection{Breakup}
To begin we will study the solutions to generalized KdV equations close 
to the breakup of the corresponding dispersionless equation. 
A generic critical point 
$(x_{c},t_{c},u_{c})$ is given by 
\begin{align}
    a(u_{c})t_{c}+\Phi(u_{c}) & =x_{c},
    \nonumber  \\
    a'(u_{c})t_{c}+\Phi'(u_{c}) &= 0,
    \label{critp}  \\
    a''(u_{c})t_{c}+\Phi''(u_{c}) &= 0,
    \nonumber
\end{align}
where $\Phi(u)$ is the inverse of the initial data  $\phi(x)$ (which might consist 
of several branches). We will always study 
the initial data $\phi(x)=\mbox{sech}^{2}x$, which imply $\Phi(u) = \ln 
((1\pm\sqrt{1-u})/\sqrt{u})$. For the critical values we obtain
\begin{align}
    u_{c}& =\frac{2n}{2n+1},
    \nonumber  \\
    t_{c} & = \frac{(1+2n)^{n+1/2}}{6(2n)^{n+1}},
    \label{critv}  \\
    x_{c} & =\frac{\sqrt{2n+1}}{2n}+\ln\left(
    \frac{\sqrt{2n+1}+1}{\sqrt{2n}}\right),
    \nonumber\\
    k &= -\frac{1}{6}(a'''(u_{c})t_{c}+\Phi'''(u_{c})) =\frac{ (2n+1)^{9/2}}{96n^{2}}
    \nonumber.
\end{align}
We study first the difference between the numerical solution to the 
generalized KdV equation and the solution to the dispersionless 
equation, a generalized Hopf equation, 
on the whole computational domain. For values of $\epsilon=10^{-1},10^{-1.25},\ldots,10^{-3}$ 
we find that this difference  scales for $n=1$ (KdV) roughly as 
$\epsilon^{\alpha}$ with $\alpha=0.299$ (correlation coefficient 
$r= 0.99997$ in linear regression, standard deviation $\sigma_{\alpha}=0.0018$), for 
$n=3$ we have $\alpha=0.317$ ($r=0.9998$, $\sigma_{\alpha}= 0.0046$),
for 
$n=4$ we have $\alpha=0.324$ ($r=0.9998$, $\sigma_{\alpha}= 0.005$),
and for 
$n=5$ we have $\alpha=0.325$ ($r=0.9998$, $\sigma_{\alpha}= 0.0053$).
The predicted value is $2/7=0.2857$. It can be seen that the above 
values are all higher, and that the scaling for the generalized KdV 
equations is close to $\epsilon^{1/3}$. Thus the decrease is at least 
of the predicted order. It is not surpising that higher values for 
the exponent are found since we consider considerably large values of 
$\epsilon$ for which the contributions of higher order in the 
difference still play a considerable role.

\begin{figure}[!htb]
\centering
\includegraphics[width=\textwidth]{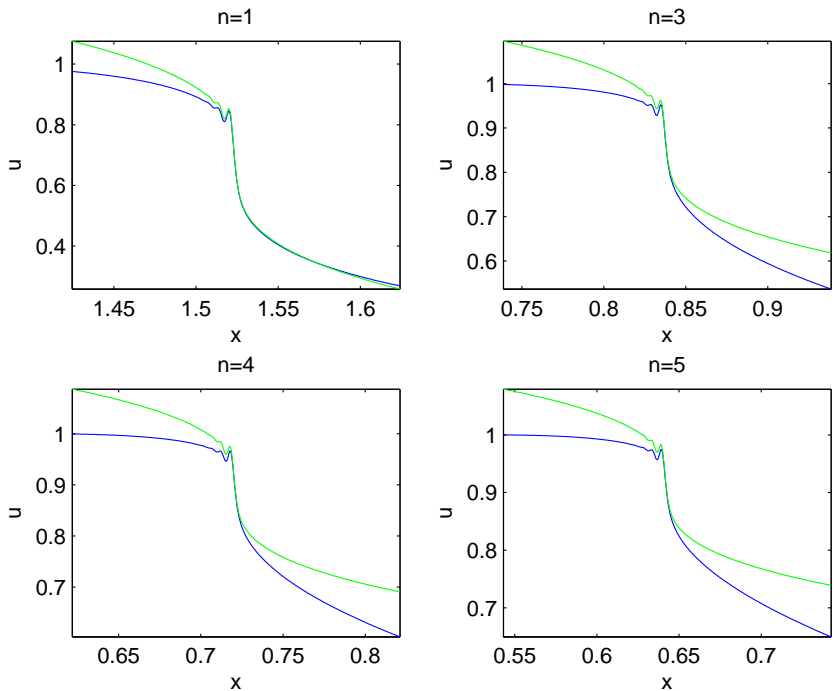}
\caption{The blue line
is the solution of the generalized KdV equation $u_t+6u^nu_x+\epsilon^2u_{xxx}=0$ for different values of $n$ for the 
initial data $\phi(x)=1/\cosh^2x$ and $\epsilon=10^{-3}$ at the time 
$t_{c}$  and near the point of 
gradient catastrophe  $x_c$ of  the Hopf solution (center of the 
figure). 
The green line is the 
multiscale approximation in terms of the PI2 solution.
}
\label{kdvg4ne3}
\end{figure}

\begin{figure}[!htb]
\centering
\begin{minipage}[c]{.45\textwidth}
\includegraphics[width=1.1\textwidth]{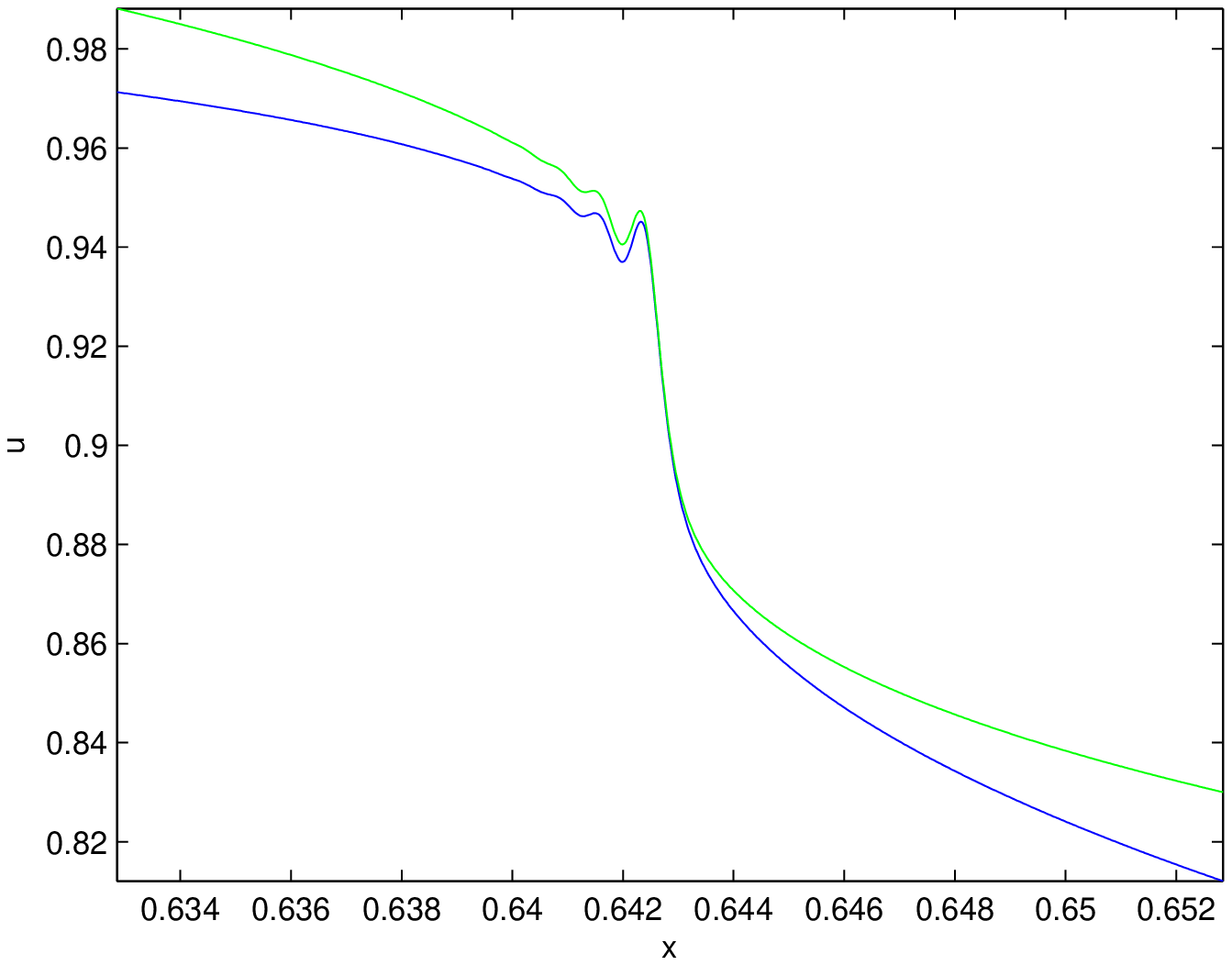}
\end{minipage}%
\hspace{5mm}%
\begin{minipage}[c]{.450\textwidth}
\includegraphics[width=1.1\textwidth]{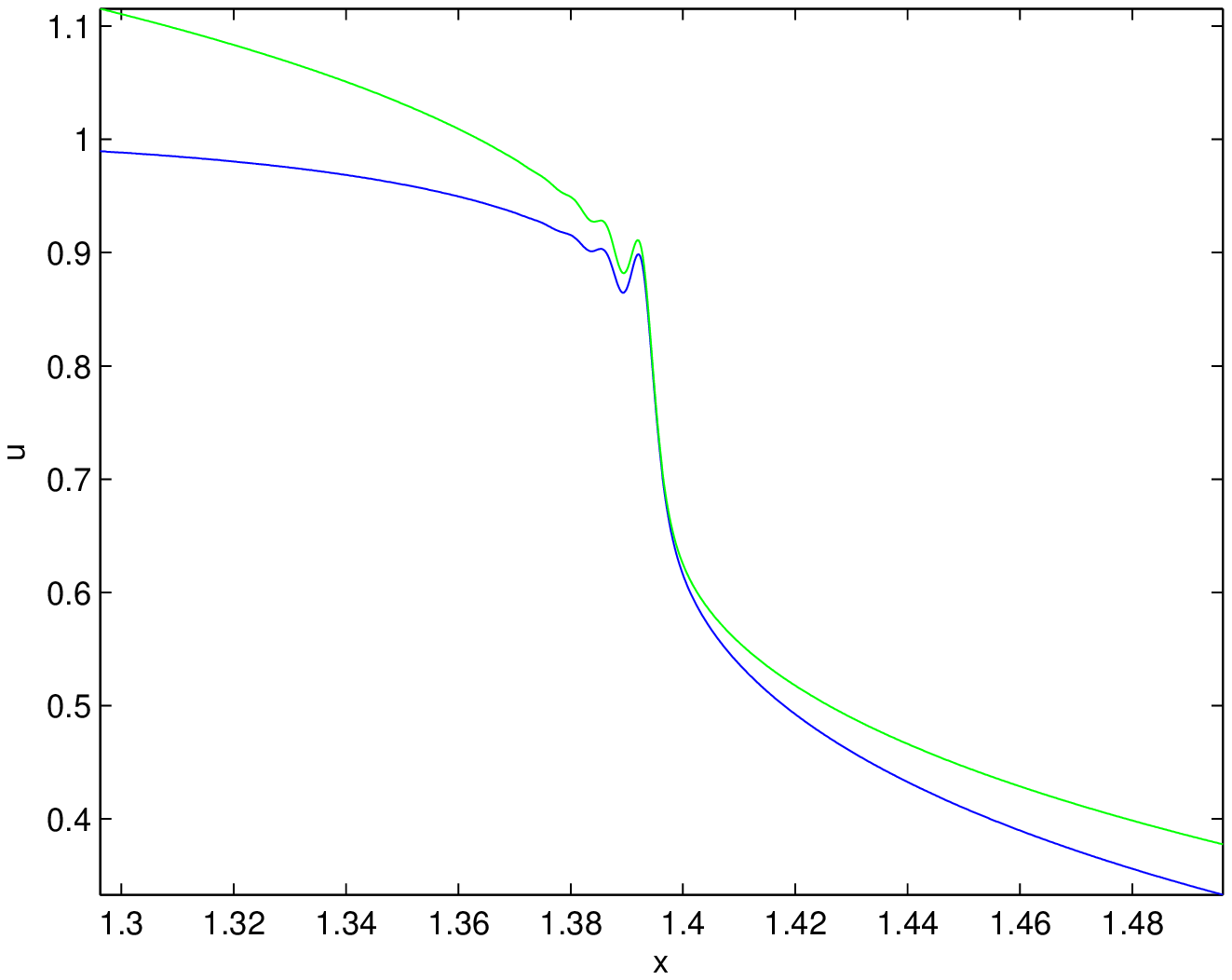}
\end{minipage}
\caption{The blue line
is the solution of $u_t+a(u)u_x+\epsilon^2u_{xxx}=0$  with $a(u)=6u^{5}$  on the left and $a(u)=6 \sinh u$ on the right, for the 
initial data $\phi(x)=1/\cosh^2x$ and $\epsilon=10^{-4}$ at the time 
$t_{c}$ near the point of 
gradient catastrophe  $x_c$ of  the Hopf solution (center of each 
figure). The green line is the 
multiscale approximation in terms of the PI2 solution.\label{kdvg5c1e8}}
\end{figure}

As discussed in the previous section it is conjectured that the behavior of the solutions to the 
generalized KdV equation in the vicinity of the critical point is 
given in terms of the special solution to the PI2 equation. Expanding 
$a(u)$ for $u\sim u_{c}$ as in \cite{dub06}, one finds the behavior 
shown in Fig.~\ref{kdvg4ne3} for different values of $n$. It can be 
seen that the asymptotic description is much better  for 
KdV due the fact that the PI2 transcendent gives an exact solution to 
KdV.  For other values of $n$ this transcendent gives the 
conjectured description in the vicinity of the critical point.
The quality of the asymptotic description shows the expected scaling for 
smaller values of $\epsilon$ as can be seen on the left in Fig.~\ref{kdvg5c1e8}.  

The quality of this PI2 approximation is not limited to 
functions $a(u)$ in (\ref{genKdV}) which are polynomial in $u$. If we 
consider the case $a(u)=6\sinh u$, we obtain the right figure in Fig.~\ref{kdvg5c1e8}.   It 
can be seen that the  PI2 asymptotics gives the same 
excellent description as for KdV.
%
\subsection{Oscillatory regimes and blowup}
It is known that solutions to initial value problems with 
sufficiently smooth initial data for the 
generalized KdV equation with $n<4$ are globally regular in time. 
This is not the case for for $n\geq4$ where blowup can occur at 
finite time with $n=4$ being the critical case. For this case a 
theorem by Martel and Merle \cite{MM} states that solutions on the 
real line, with negative energy, blow up in finite or infinite time. 
For the general case $n>4$ and periodic settings considered here,
the question is still open. Since the energy has the form
$$E = 
\int_{\mathbb{R}}^{}\left[\frac{\epsilon^2}{2}u_{x}^{2}-\frac{6u^{n+2}}{(n+1)(n+2)}\right]\, dx$$
it will be always negative for sufficiently small $\epsilon$ and 
positive $u$. 

\begin{figure}[!htb]
\centering
\includegraphics[width=0.45\textwidth]{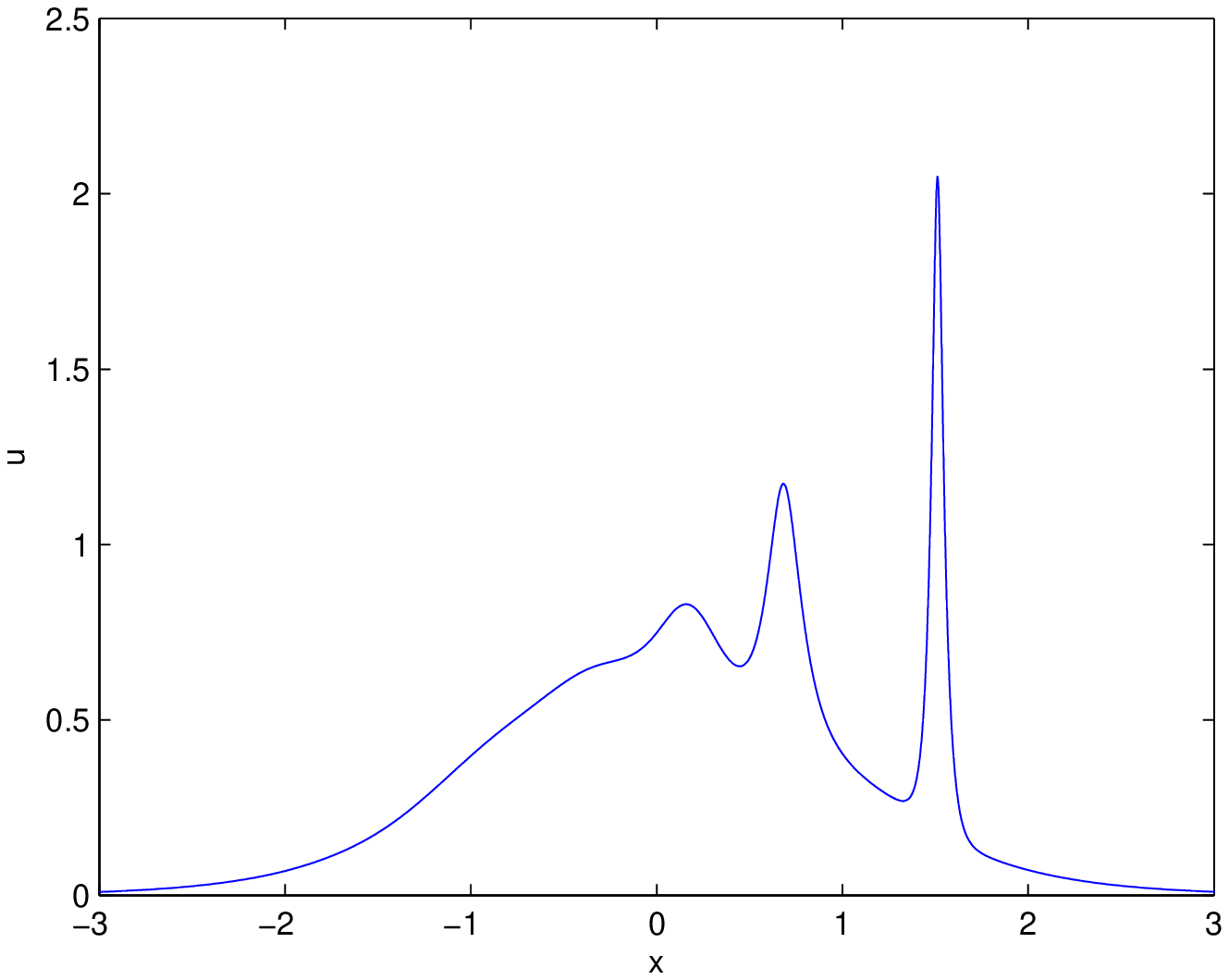}
\includegraphics[width=0.45\textwidth]{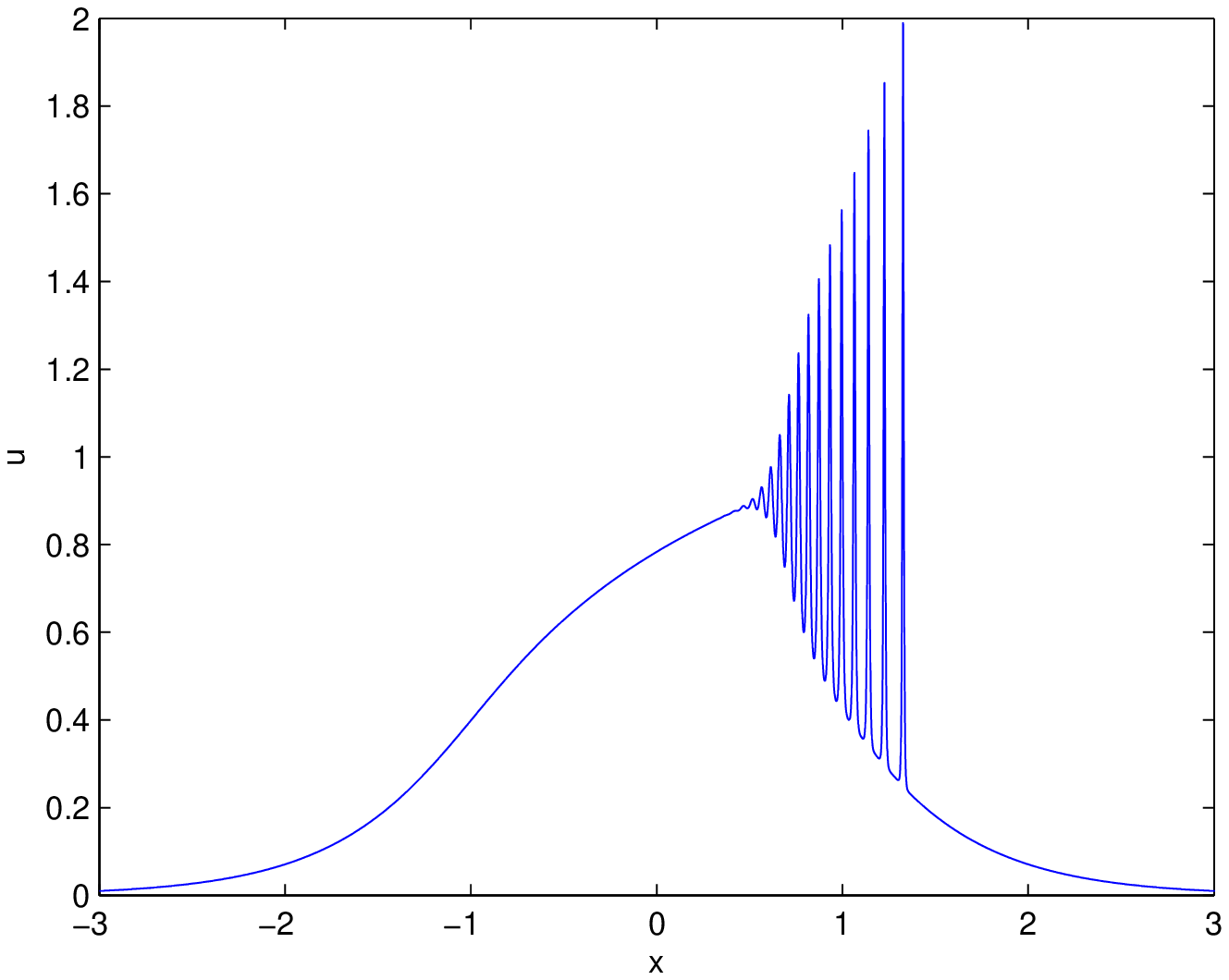}
\caption{Solution of the generalized KdV equation with $n=4$ for the 
initial data $\phi(x)=1/\cosh^2x$ and $\epsilon=10^{-1}$ at the time 
$t=0.3180\gg t_{c}$ on the left, and for $\epsilon=10^{-2}$ at the time 
$t=0.2235\gg t_{c}$ on the right. 
}
\label{kdvg4_e2_3180}
\end{figure}

\begin{figure}[!htb]
\centering
\begin{minipage}[c]{.45\textwidth}
\includegraphics[width=1.2\textwidth]{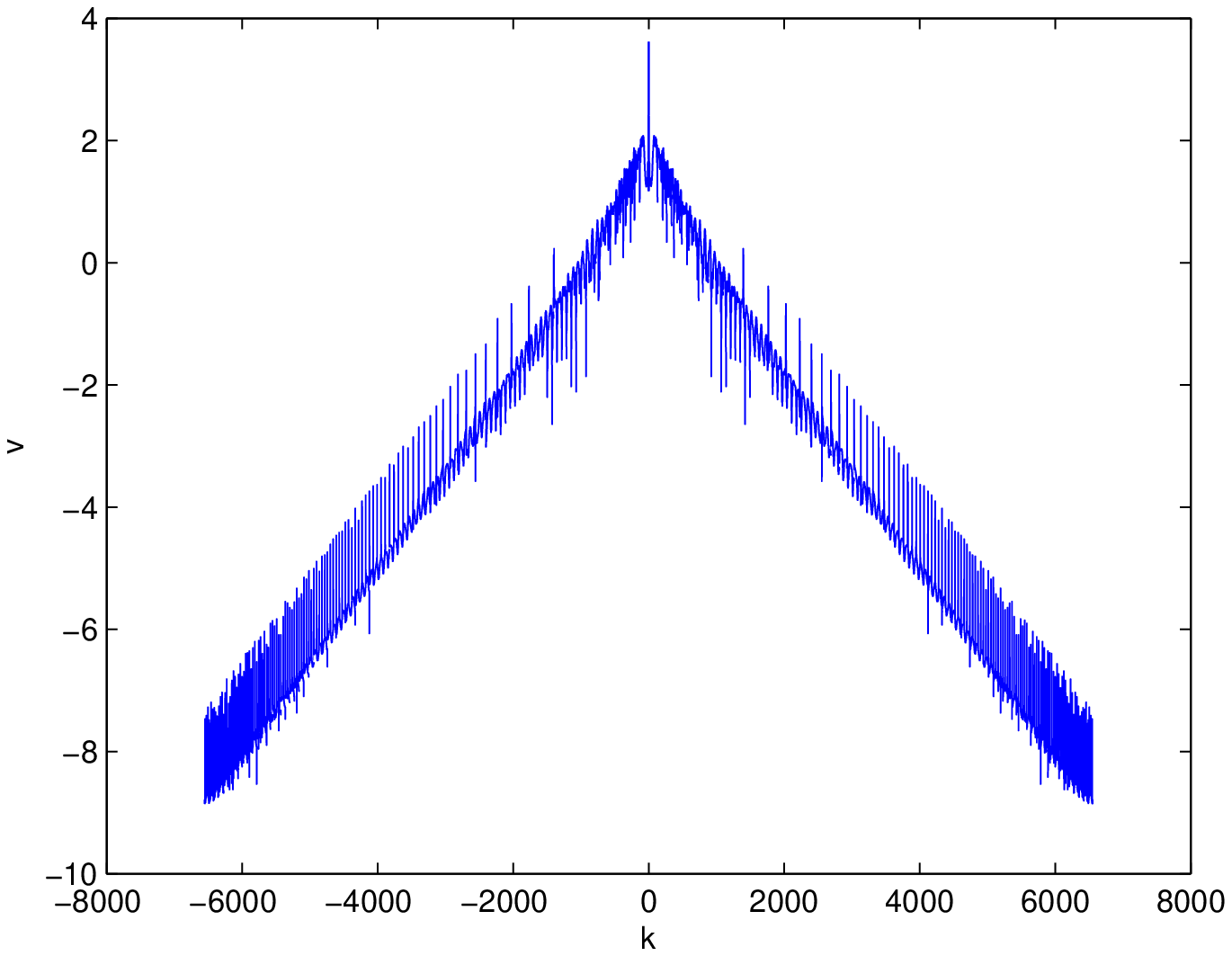}
\caption{\noindent Logarithm with base 10 of the modulus of the Fourier coefficents of the 
function shown in Fig.~\ref{kdvg4_e2_3180} on the right. 
}
\label{kdvg4_e4_2235f}
\end{minipage}%
\hspace{5mm}%
\begin{minipage}[c]{.450\textwidth}
\includegraphics[width=1.3\textwidth]{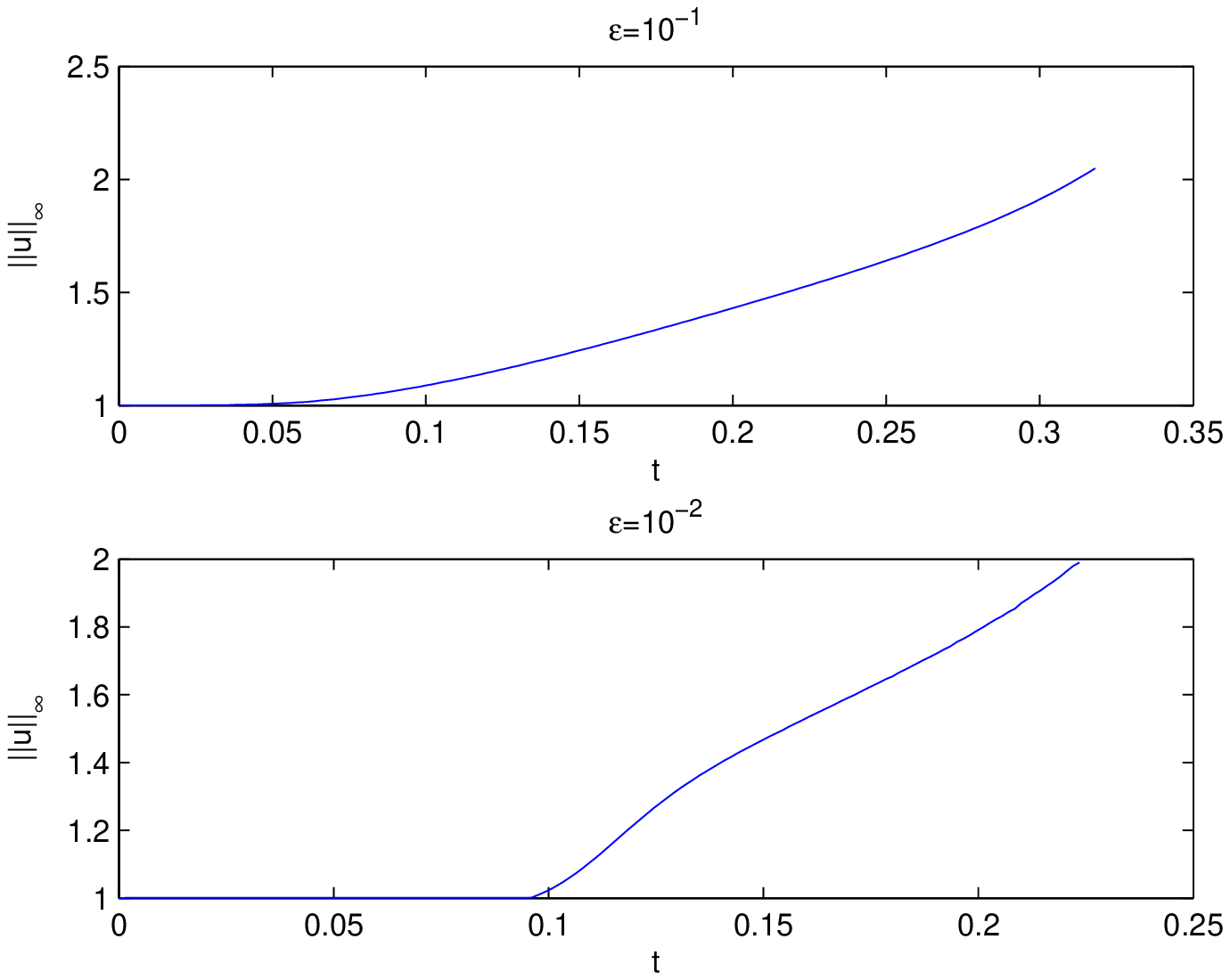}
\caption{$L^{\infty}$ norm of the solutions in 
Fig.~\ref{kdvg4_e2_3180}. 
}
\label{kdvg4_infty_2e}
\end{minipage}
\end{figure}

Here we address numerically the question whether the formation of 
dispersive shocks, i.e., of a region of rapid modulated oscillations, 
precedes a potential blowup. We expect that the breakup of the 
solution to the dispersionless equation is regularized by the 
dispersion in the form of oscillations which then develop into blowup 
if the latter exists. This is exactly what we see in the following. 
Notice that the breakup time is given by the dispersionless equation and 
is thus independent of $\epsilon$.
We first study the case $n=4$. For $\epsilon=1$, the energy is 
positive and no indication of blowup is observed. 
For $\epsilon=0.1$ we obtain for the 
initial data  $\phi(x)=\mbox{sech}^{2}x$ the left figure in 
Fig.~\ref{kdvg4_e2_3180}. 
For smaller $\epsilon$ ($\epsilon=0.01$) the behavior is similar, but there are as 
expected more oscillations, and the size of the 
oscillations reaches higher values earlier as can be seen in 
Fig.~\ref{kdvg4_e2_3180}. 
For obvious reasons it is numerically difficult to decide whether the 
dispersive shock will lead to a blowup. In practice we run out of 
resolution before the code breaks down because of a blowup. This is 
due to oscillations in Fourier space as can be 
seen in Fig.~\ref{kdvg4_e4_2235f}. Though there is in principal enough 
resolution to approach $u(x,t)$, the oscillations of  the Fourier 
coefficients make an accurate approximation via a Fourier transform 
impossible. 

The reason for this behavior is as discussed in 
\cite{SS82} that singularities of the form $(z-z_{j})^{\mu_{j}}$ in 
the complex plane lead asymptotically to Fourier coefficients with 
modulus of the 
form $Ck^{-(\mu_{j}+1)}\exp(-\delta k)$, $\delta>0$. If there are several such 
singularities, there will be oscillations in the Fourier 
coefficients. In the present case there are at least two such 
singularities, the breakup which is strictly speaking only singular 
for $\epsilon=0$, but has an effect already for finite $\epsilon$, 
and the blowup, which leads to the behavior seen in 
(\ref{kdvg4_e4_2235f}). 
In  Fig.~\ref{kdvg4_infty_2e} we give the $L^{\infty}$-norm of the 
solutions in Fig.~\ref{kdvg4_e2_3180}. 
It cannot be decided on the base of these numerical data whether 
there is finite time blowup in this case. If it exists it is clearly 
preceded by a dispersive shock. 

\begin{figure}[!htb]
\centering
\includegraphics[width=0.45\textwidth]{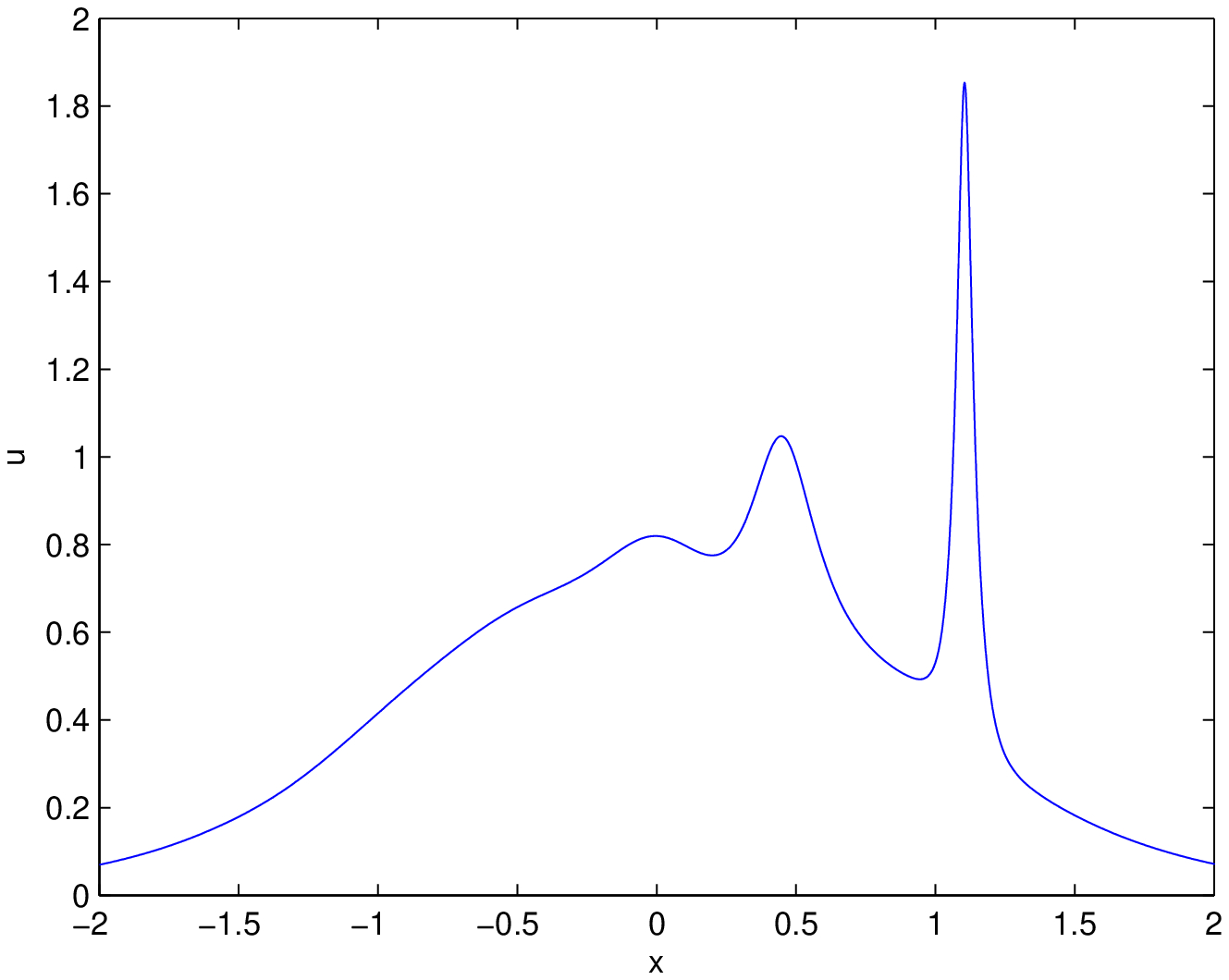}
\includegraphics[width=0.45\textwidth]{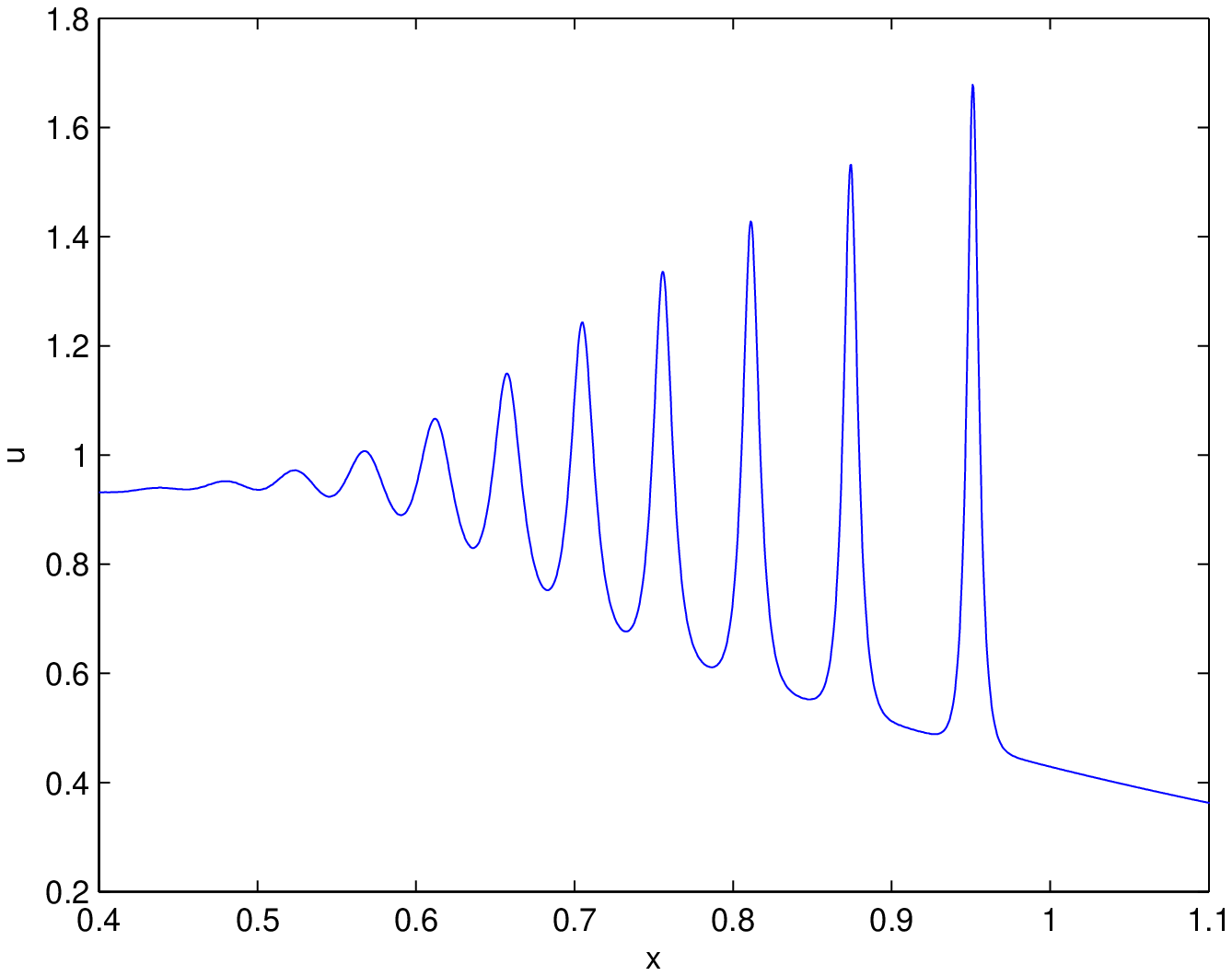}
\caption{Solution of the generalized KdV equation with $n=5$ for the 
initial data $\phi(x)=1/\cosh^2x$ and $\epsilon=10^{-1}$ at the time 
$t=0.2362\gg t_{c}$ on the left and for $\epsilon=10^{-2}$ at the time 
$t=0.2235\gg t_{c}$ on the right.  
}
\label{kdvg5_e2_2362}
\end{figure}

In the supercritical case $n=5$ we obtain a similar picture. In 
Fig.~\ref{kdvg5_e2_2362} we see the solution in the case 
$\epsilon=0.1$. Again it appears as if the rightmost peak evolves 
into a singularity.
For smaller $\epsilon$ ($\epsilon=0.01$)  there are again more 
oscillations, which stresses the importance of dispersive 
regularization before a potential blowup. 
Studying the  $L^{\infty}$-norm of the 
solutions in Fig.~\ref{kdvg5_e2_2362}, we can 
see that the case $\epsilon=0.1$ indeed seems to approach an 
$L^{\infty}$ blowup in finite time. Because of resolution problems we could not 
reach a similar point for $\epsilon=0.01$.  
\begin{figure}[!htb]
\centering
\includegraphics[width=0.7\textwidth]{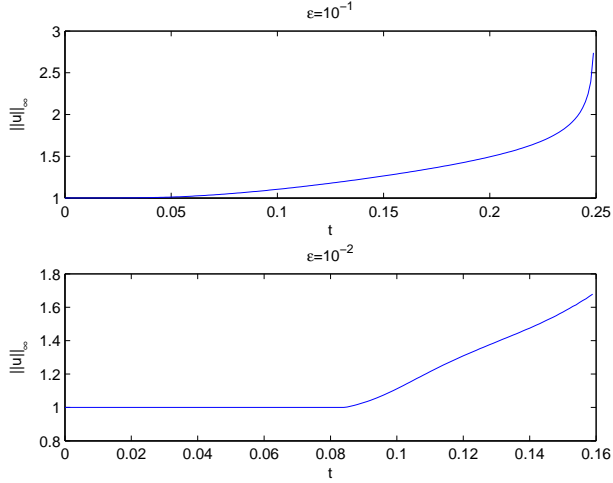}
\caption{$L^{\infty}$ norm of the solutions in 
Fig.~\ref{kdvg5_e2_2362}. 
}
\label{kdvg5_infty_2e}
\end{figure}

\section{Kawahara equations}
The Kawahara equations  \cite{kawa} which appear in general dispersive media where 
the effects of the third order derivative is weak as in certain hydrodynamic 
or magneto-hydrodynamic settings can be written in the form
\begin{equation}
    u_{t}+\frac{1}{2}\partial_{x}f(u,\epsilon\, u_{x},\epsilon^2 u_{xx})
    +\beta\, \epsilon^4 u_{xxxxx} = 0.
    \label{kawa}
\end{equation}
Here we will mainly study the case 
\beq\label{kawa1}
f(u,\epsilon\, u_{x},\epsilon^2 
u_{xx})=6u^{2} +2\alpha\,\epsilon^2  u_{xx}
\eeq 
with $\alpha=1$ and $\beta=\pm1$. 
The global well posedness of  solutions of (\ref{kawa}) in a suitable Sobolev space has been proved in \cite{ponce}. 

The functional parameters $c(u)$, $p(u)$  in (\ref{cub2}) are constants
\beq\label{kawa-cp}
c(u)=\frac16\alpha, \quad p(u)=\frac1{12} \beta.
\eeq

\begin{figure}[!htb]
\centering
\begin{minipage}[c]{.45\textwidth}
\includegraphics[width=1.2\textwidth]{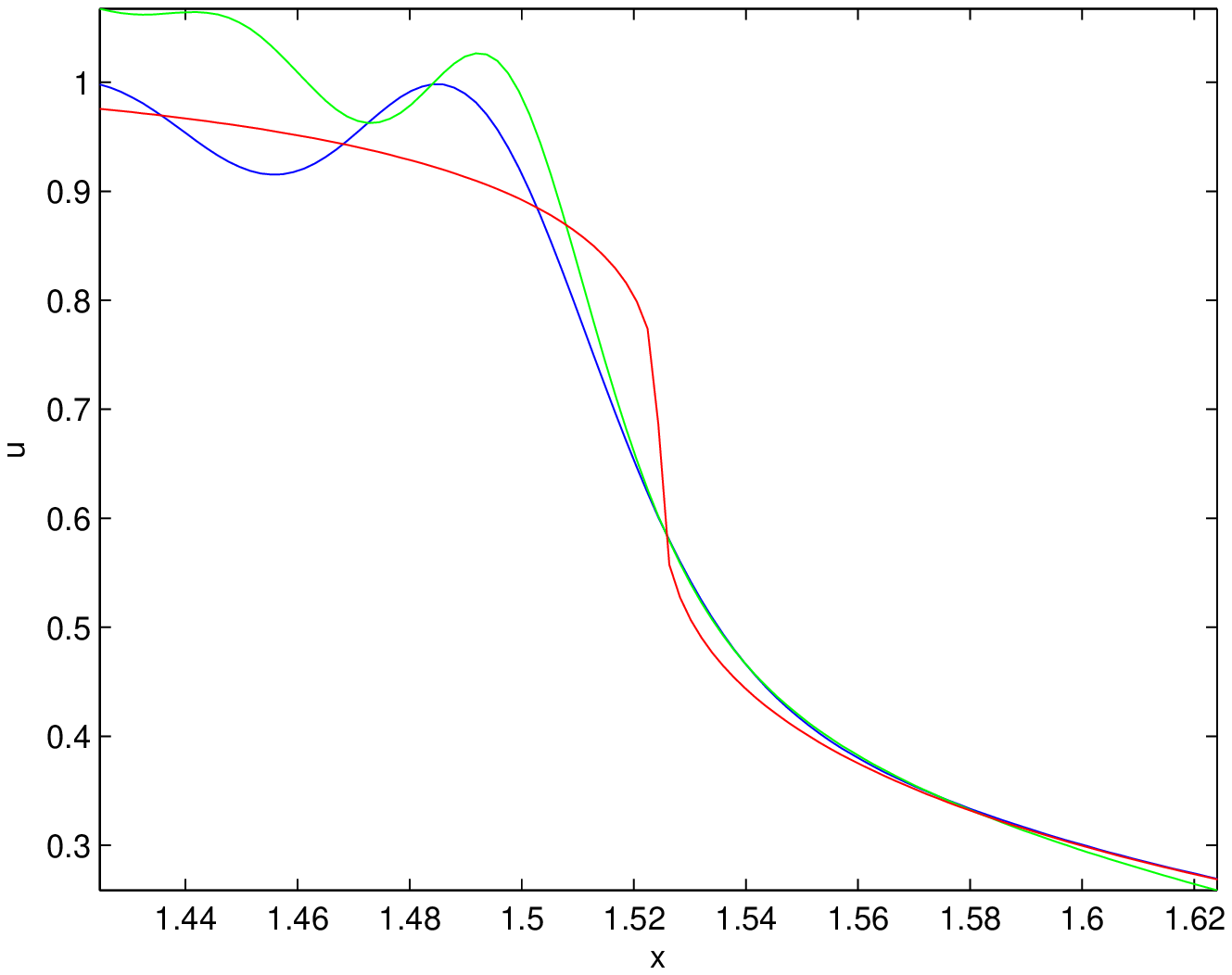}
\end{minipage}%
\hspace{5mm}%
\begin{minipage}[c]{.450\textwidth}
\includegraphics[width=1.2\textwidth]{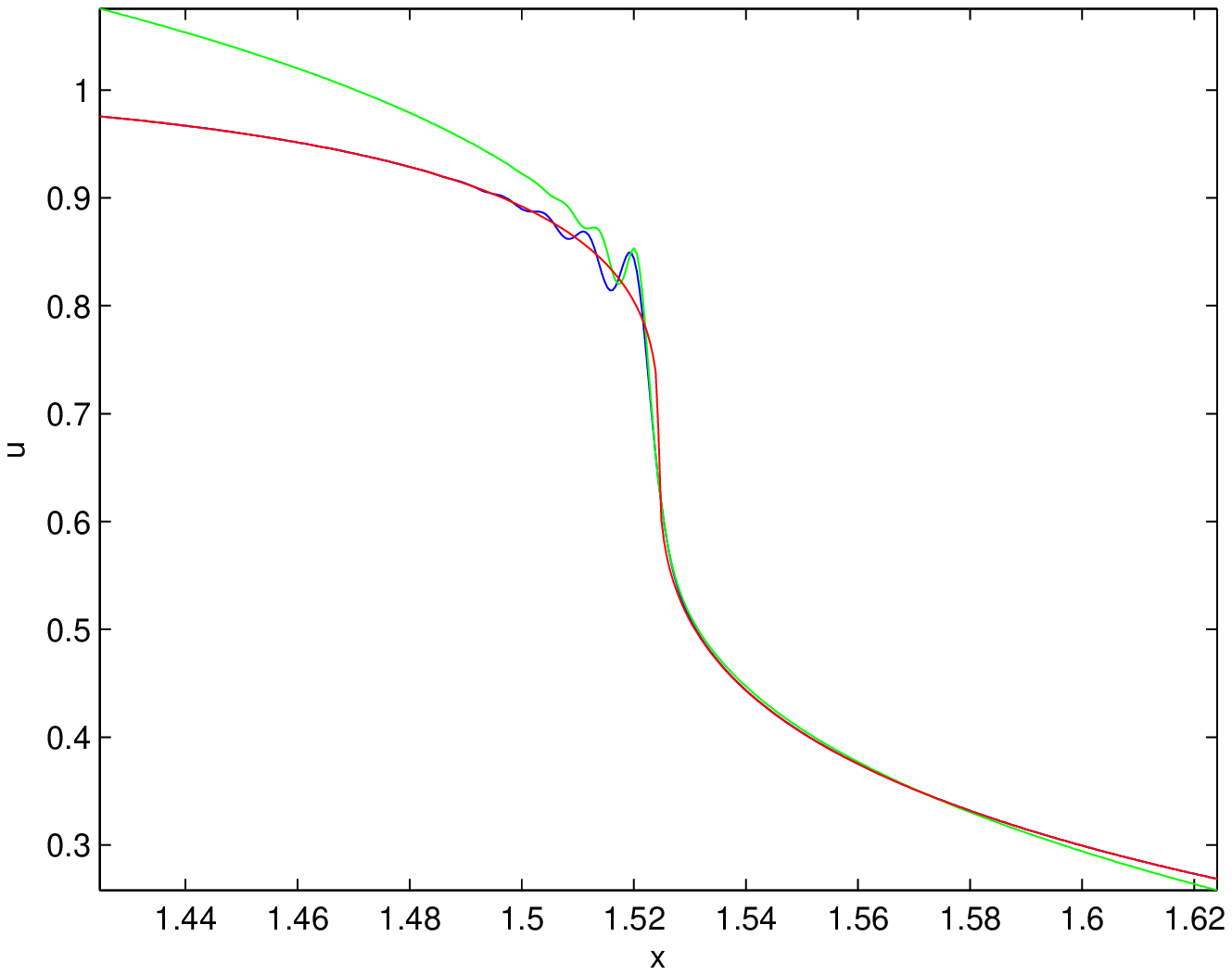}
\end{minipage}
\caption{The blue line
is the solution of the Kawahara equation ($\beta=-1$) for the 
initial data $\phi(x)=1/\cosh^2x$ and $\epsilon=10^{-2}$ on the left  figure and $\epsilon=10^{-3}$ on the right figure. The plot is taken  at the time 
$t_{c}$ near the point of 
gradient catastrophe  $x_c$ of  the Hopf solution (center of the 
figure).  Here  $x_c\simeq 1.524 $, $t_c\simeq 0.216$.
The red line is the corresponding  Hopf solution, the green line the 
multiscale approximation in terms of the PI2 solution.
}
\label{kawa1e4.0cm}
\end{figure}

At the critical point we obtain for $\beta=-1$ that the breakup 
behavior is well described by PI2 in lowest order as can be seen in 
Fig.~\ref{kawa1e4.0cm}.

It can be seen that the PI2 solution gives close to the breakup point 
a much better description of the Kawahara solution than the 
corresponding Hopf solution. The oscillation closest to the breakup 
point is too far away from the latter to be correctly reproduced, but 
the PI2 solution catches qualitatively the oscillatory behavior of 
the Kawahara solution near the critical point. With smaller 
$\epsilon$, the agreement gets as expected better, see the right figure in 
Fig.~\ref{kawa1e4.0cm}.
%

\begin{figure}[!htb]
\centering
\begin{minipage}[c]{.45\textwidth}
\includegraphics[width=1.2\textwidth]{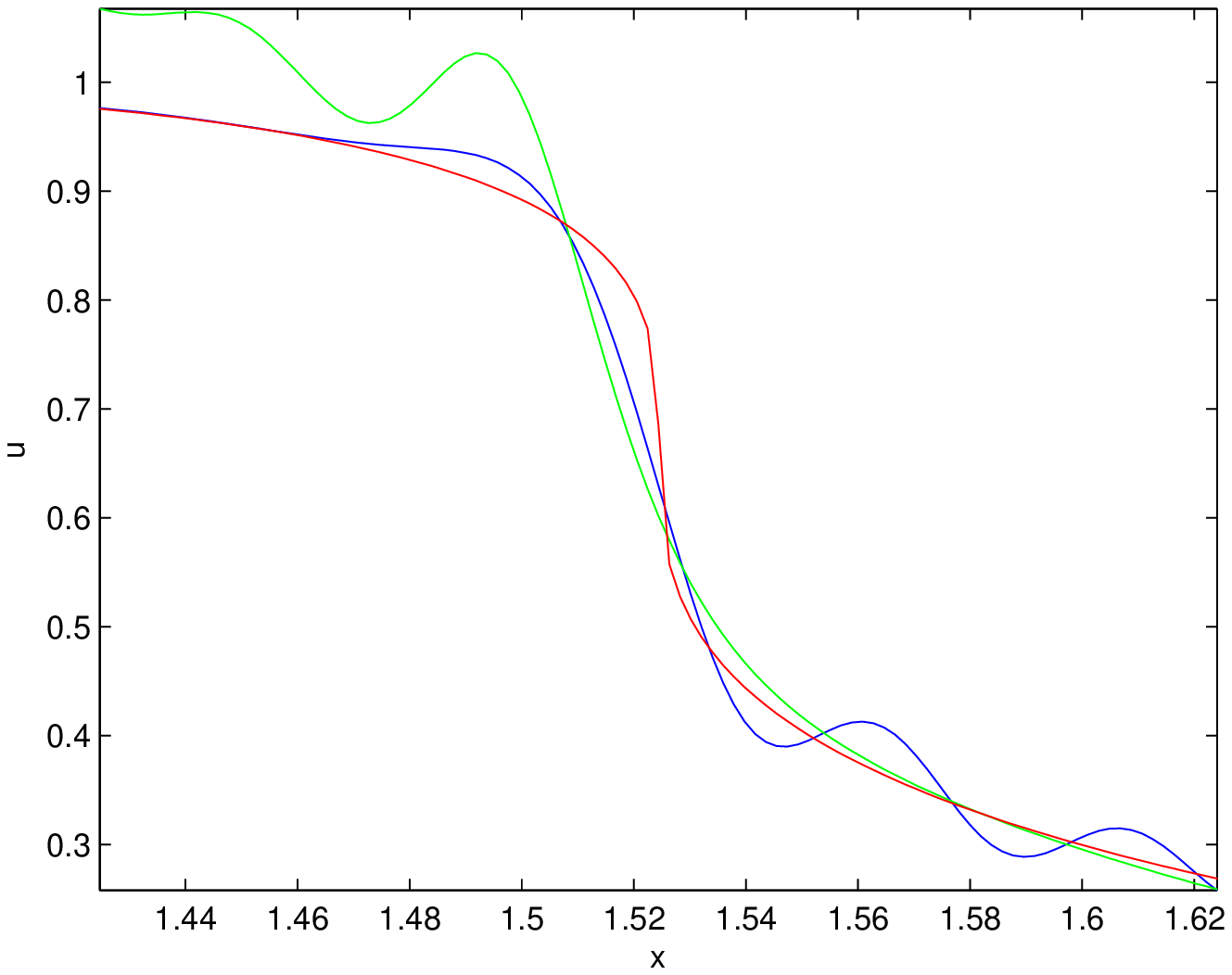}
\end{minipage}%
\hspace{5mm}%
\begin{minipage}[c]{.450\textwidth}
\includegraphics[width=1.2\textwidth]{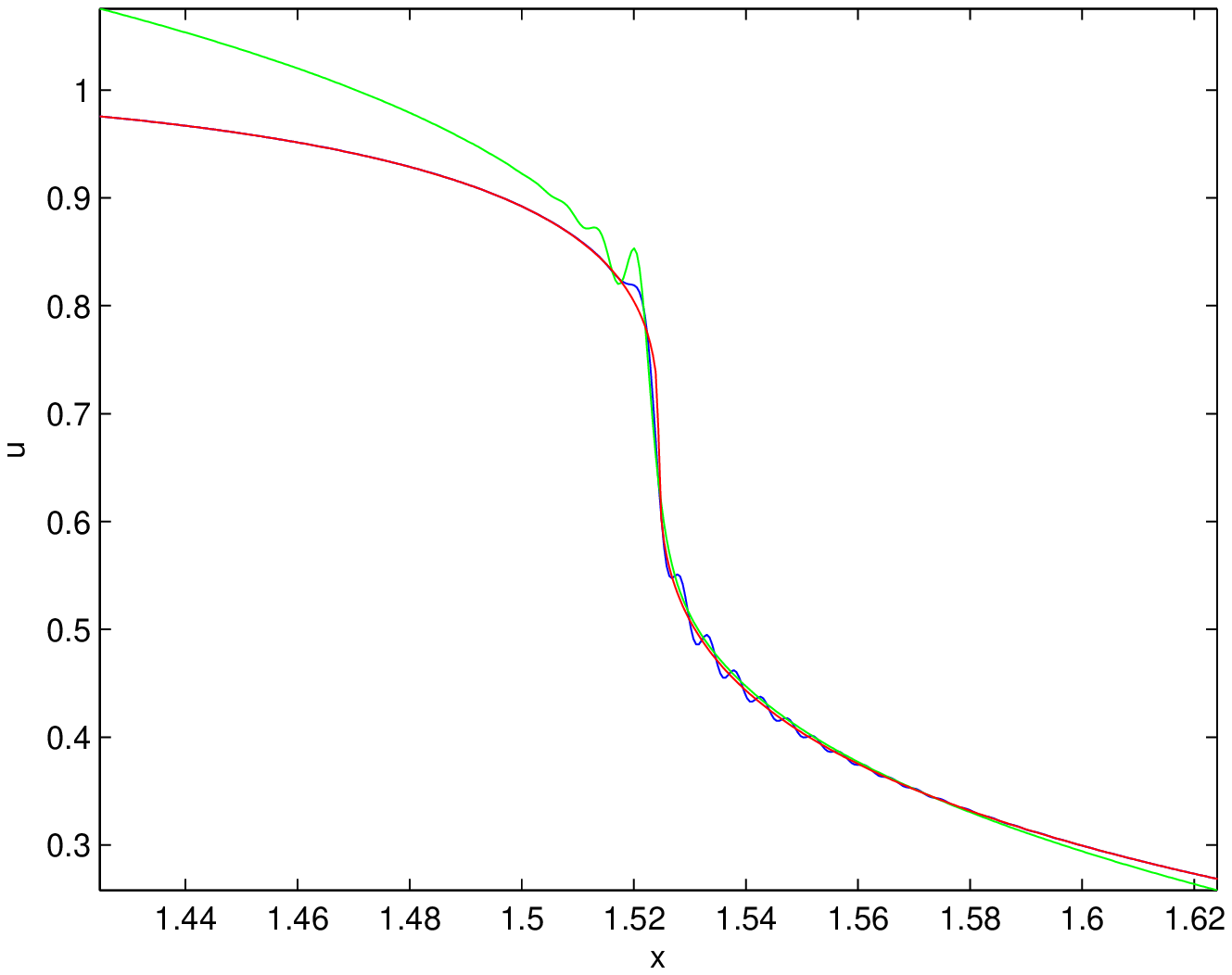}
\end{minipage}
\caption{The blue line
is the solution of the Kawahara equation ($\beta=1$) for the 
initial data $\phi(x)=1/\cosh^2x$ and $\epsilon=10^{-2}$ on the left  figure and $\epsilon=10^{-3}$ on the right figure. The plot is taken  at the time 
$t_{c}$ near the point of 
gradient catastrophe  $x_c$ of  the Hopf solution (center of the 
figure).  Here  $x_c\simeq 1.524 $, $t_c\simeq 0.216$.
The red line is the corresponding  Hopf solution, the green line is  the 
multiscale approximation in terms of the PI2 solution.
}
\label{kawa1e4.0c}
\end{figure}

For $\beta=1$, the breakup behavior of solutions to the Kawahara 
changes as can be seen from Fig.~\ref{kawa1e4.0c}.
In this case the oscillations in the Kawahara solution appear on the 
other side of the critical point and around it with small amplitude. 
This behaviour cannot be captured by the PI2 solution, but it is a 
higher order effect. Close to the critical point, the multiscale 
solution gives as before a much better description of the Kawahara 
solution than the Hopf solution. 

For smaller values of $\epsilon$, both asymptotic solutions become 
more satisfactory as can be seen from  the right figure in   Fig.~\ref{kawa1e4.0c}.
\begin{figure}[!htb]
\centering
\includegraphics[width=.7\textwidth]{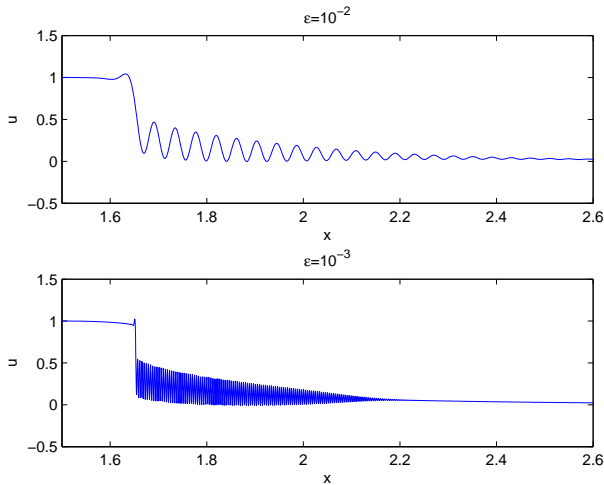}
\caption{Oscillatory zone of the solutions to the Kawahara equation ($\beta=1$) for the 
initial data $\phi(x)=1/\cosh^2x$ and two values of $\epsilon$ at time 
$t=0.25>t_{c}$.
}
\label{kawa2e_.25}
\end{figure}
It is interesting to notice that with decreasing $\epsilon$, the 
oscillations become smaller in amplitude in this case, but appear 
closer to the critical point. It can also be seen that the solution 
has the tendency to form one oscillation on the other side of the 
critical point close to the corresponding PI2 oscillation. Tracing 
the solution for larger times, it can be recognized that this will be 
the only oscillation to the left of the critical point, whereas a 
zone of high-frequent oscillations which appears to be essentially 
unbounded (see \cite{hunter})  develops to the right, see Fig.~\ref{kawa2e_.25}. The 
oscillations appear to be as in the KdV case more and more confined 
to a zone similar to the Whitham zone, though no asymptotic 
description of the oscillations exists since the equation is not 
integrable.

\begin{figure}[!htb]
\centering
\includegraphics[width=.6\textwidth]{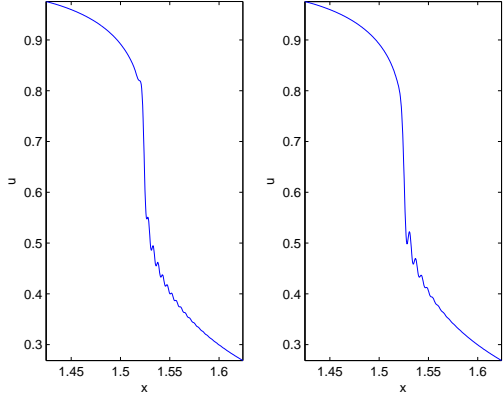}
\caption{Solutions of the Kawahara equation~\ref{kawa}  with $\beta=1$  $\alpha=1$ to the left and $\alpha=0$ to the right. The  solution is given for the 
initial data  $\phi(x)=1/\cosh^2x$ and $\epsilon=10^{-3}$ at the time 
$t_{c}$ near the point of 
gradient catastrophe  $x_c$ of  the Hopf solution. }
\label{kawac2a}
\end{figure}
It seems also that this one oscillation to the left is really due to 
the third order derivative in the Kawahara equation as can be seen in 
Fig.~\ref{kawac2a} where one has to the left the Kawahara solution 
from Fig.~\ref{kawa1e4.0c} and to the right the analogous solution 
for $\alpha=0$, i.e., Kawahara without third order derivative. The 
oscillations to the right of the critical point being due to the 
fifth order derivative are present in both cases and  have only 
slightly different form.

\subsection{PDE with nonlinear dispersion}
To show that  the breakup behavior discussed in the previous sections 
is typical, we will now consider equations of  the form (\ref{ur}) 
with nonlinear dispersion, i.e., with functions $c(u)$ and $p(u)$ not 
constant. The 
Camassa-Holm equation (CH) falls in this class if the nonlocal term is 
expanded in a von Neumann series, see \cite{dub06}, for functions 
$c(u)\sim u$ and $p(u)\sim u$. The applicability of 
the PI2 asymptotics to CH was studied numerically in \cite{gk08}. 
For simplicity we restrict our analysis to the case $c(u)$ and $p(u)$ 
both proportional to $u^{2}$ with the Hopf equation as the 
dispersionless equation and initial data of the form 
$\phi(x)=\mbox{sech}^{2}x$. More complicated functions $c$ and $p$ can be 
considered, but the results are qualitatively the same. 

\begin{figure}[!htb]
\centering
\begin{minipage}[c]{.45\textwidth}
\includegraphics[width=1.2\textwidth]{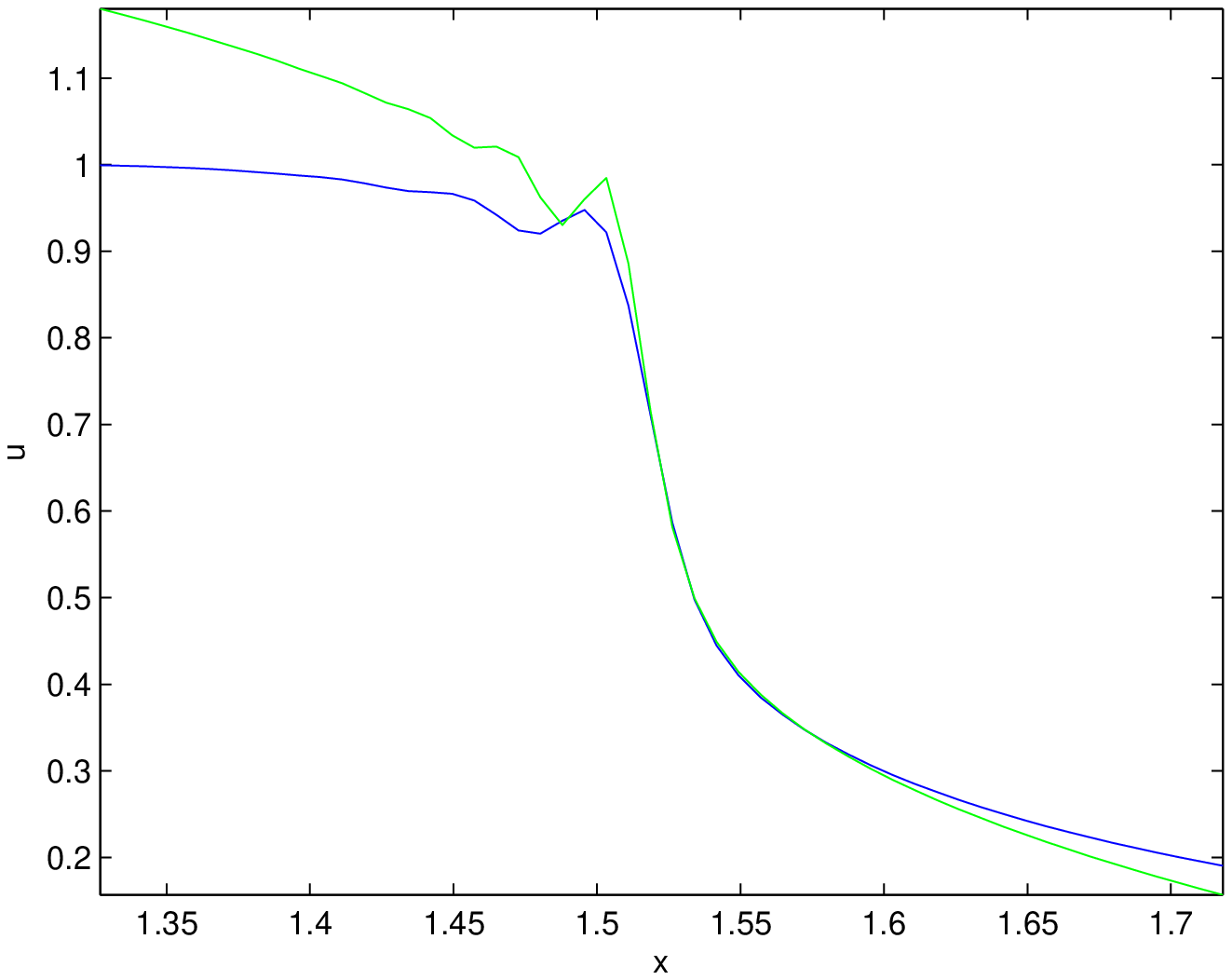}
\caption{Solution to the equation (\ref{ur}) for $a(u)=u$, $c(u)=u^{2}$ and 
$p(u)=0$ and initial data $\phi(x)=\mbox{sech}^{2}x$ at the critical 
time, 
and the corresponding multiscale solution in terms of the PI2 
transcendent.
}
\label{dubc1c}
\end{minipage}%
\hspace{5mm}%
\begin{minipage}[c]{.450\textwidth}
\includegraphics[width=1.2\textwidth]{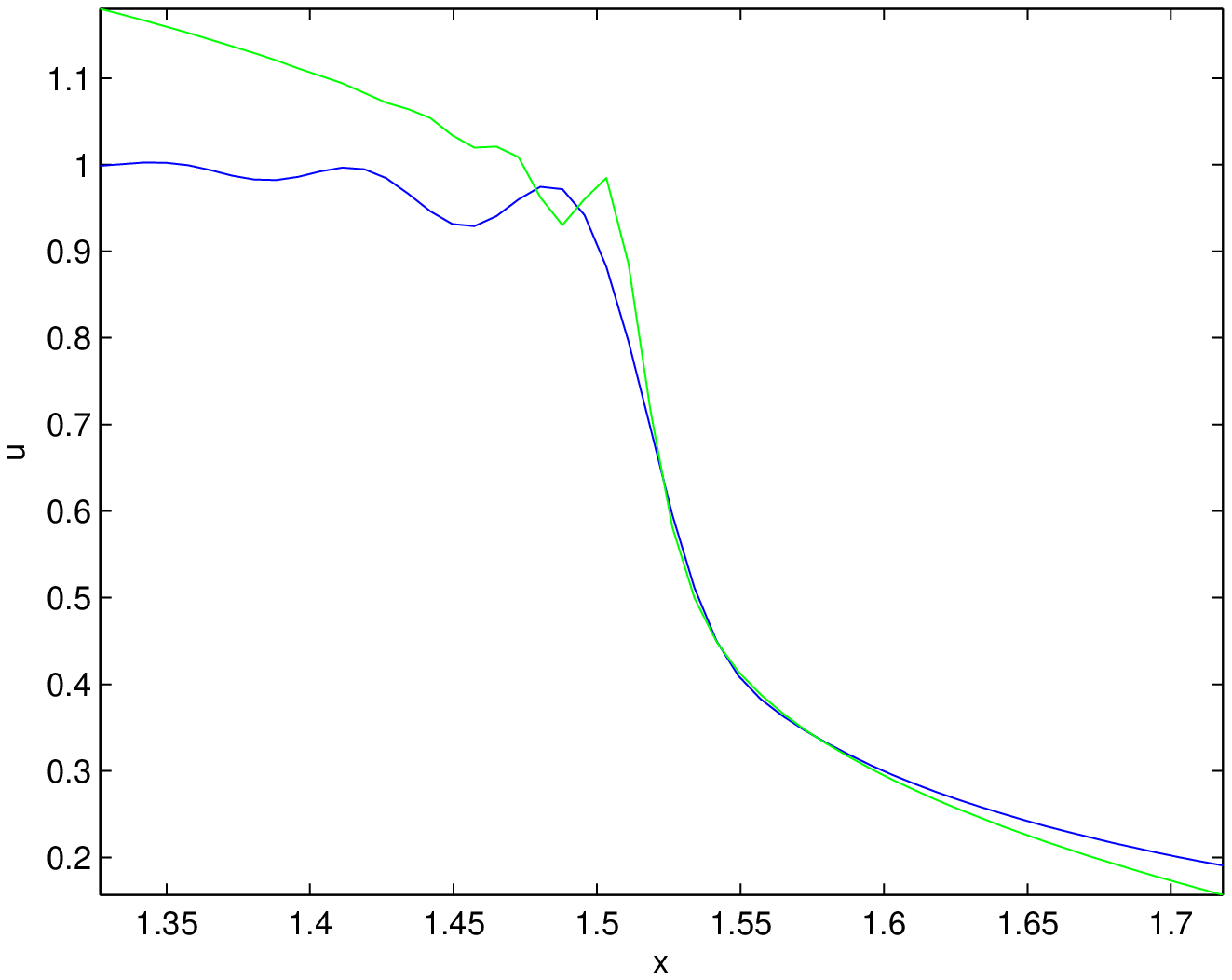}
\caption{Solution to the equation (\ref{ur}) for $a(u)=u$, 
$c(u)=-p(u)=u^{2}$
and initial data $\phi(x)=\mbox{sech}^{2}x$ at the critical time, 
and the corresponding multiscale solution in terms of the PI2 
transcendent.
}
\label{dubc1p1c}
\end{minipage}
\end{figure}

In Fig.~\ref{dubc1c} the behavior at the critical time can be seen 
for $c(u)=u^{2}$ and $p(u)=0$. The situation is obviously as in the KdV case.

\begin{figure}[!htb]
\centering
\includegraphics[width=0.7\textwidth]{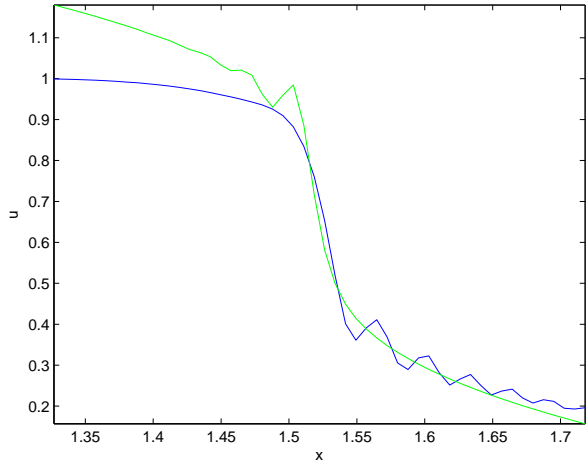}
\caption{Solution to the equation (\ref{ur}) for $a(u)=u$, 
$c(u)=p(u)=u^{2}$
and initial data $\phi(x)=\mbox{sech}^{2}x$ at the critical point, 
and the corresponding multiscale solution in terms of the PI2 
transcendent.
}
\label{dubc1pm1c}
\end{figure}

As for the Kawahara equation the relative sign between the third and 
the fifth derivative is important for the form of the oscillations. 
The situation with the opposite sign of $c$ and $p$ can be seen in 
Fig.~\ref{dubc1p1c}. It is qualitatively the same as in the KdV case.
New features appear as in the case of the Kawahara equation for the 
same sign in front of the third and fifth 
derivative. As can be seen in Fig.~\ref{dubc1pm1c}, oscillations of 
small amplitude appear as in the Kawahara equation on the other side 
of the critical point.
Thus non constant functions $c(u)$ and $p(u)$ as expected 
do not change the picture from the case of constant functions as 
long as they do not vanish at the critical point.

\subsection{Quasi-trivial transformation}
\begin{figure}[!htb]
\hskip -0.5cm
\includegraphics[width=.7\textwidth]{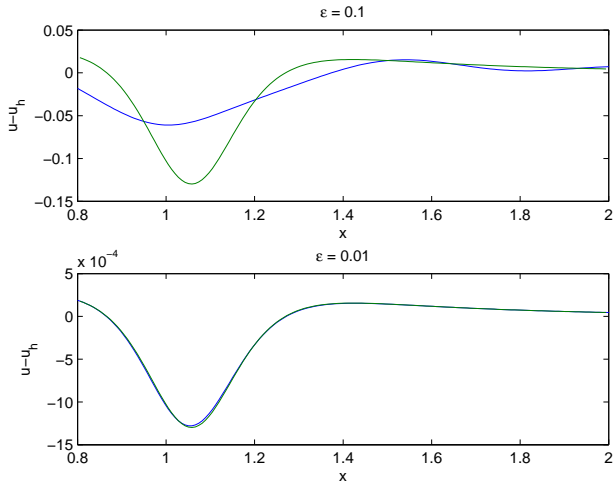}
\caption{Solution to the Kawahara equation with $\alpha=1$ for the 
initial data $\phi(x)=1/\cosh^{2}x$ at the time $t=t_{c}/2$ for two 
values of $\epsilon$; in blue 
the difference between the Kawahara and the corresponding Hopf 
solution, in green the order $\epsilon^{2}$ term of the 
quasitriviality transformation. 
}
\label{testquasi}
\end{figure}

In this subsection we study numerically the validity of the expansion 
given in  sect.~4.
For times $t\ll t_{c}$ the behavior of the solution of (\ref{eq1})  should be 
described  to order $\epsilon^{2}$ by the solution of the Hopf equation $u_t+uu_x=0$ with the same initial data. In 
fact we find that the difference between the Hopf solution and the solution to 
Kawahara equation (\ref{kawa})  with $\beta=1$, $\alpha=1$ for the initial data $\phi(x)=\mbox{sech}^{2}x$ at  
$t=t_{c}/2$ scales as $\epsilon^{\gamma}$ with $\gamma=1.94$ (values 
of $\epsilon=10^{-1}, 10^{-1.125},\ldots,10^{-3}$, correlation coefficient 
$r= 0.9997$ in linear regression, standard deviation 
$\sigma_{\alpha}=0.027$). For the same setting in the 
interval $x\in [0.8,2]$, the difference  
between the quasitriviality solution as described in sect.~4 and the 
Kawahara solution scales as $\epsilon^{\gamma}$ with $\gamma=3.77$ (correlation coefficient 
$r= 0.999$ in linear regression, standard deviation 
$\sigma_{\alpha}=0.088$). This confirms the theoretical expectations. 
In Fig.~\ref{testquasi} the difference between the Kawahara and the 
Hopf solution and the quasitriviality transform in order 
$\epsilon^{2}$ can be seen for this case.
A similar scaling is observed for $\alpha=-1$ and $\alpha=0$ (KdV). 

The difference between the solution of  generalized KdV equation $u_t+u^5u_x+\epsilon^2u_{xxx}=0$  and  the solution of the corresponding conservation law scales,  at $t=t_c/2$ as $\epsilon^{\gamma}$ with $\gamma=1.9890$ (values 
of $\epsilon=10^{-1}, 10^{-1.125},\ldots,10^{-3}$, correlation coefficient 
$r= 0.99998$ in linear regression, standard deviation 
$\sigma_{\alpha}=0.0068$).
\subsection{Second equation in the KdV hierarchy}
\begin{figure}[!htb]
\hskip -0.5cm
\includegraphics[width=1.1\textwidth]{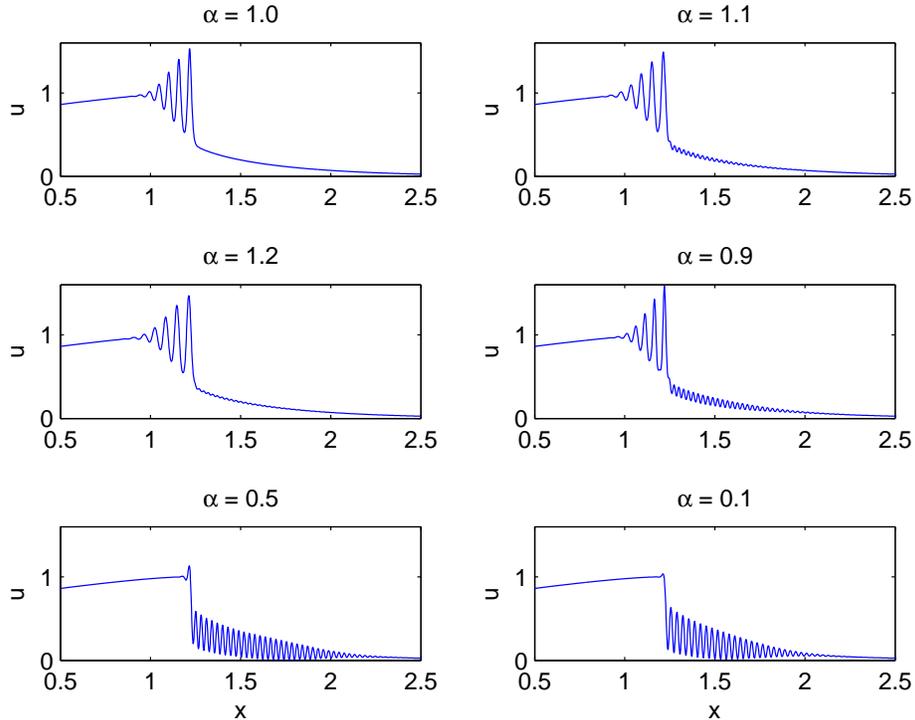}
\caption{Oscillatory part of the solution to the KdVII equation  for the 
initial data $\phi(x)=1/\cosh^2x$ and $\epsilon=10^{-2}$ at a time 
$t=0.04>t_{c}=0.029$ for several values of $\alpha$.
}
\label{kdvII4alpha}
\end{figure}

In this subsection we study the formation of dispersive shock waves for a family of equations that includes integrable and non-integrable PDEs.
Interestingly the family of equations 
\beq\label{kdv2}
u_{t}+30u^{2}u_{x}+10\alpha\epsilon^{2}(uu_{xxx}+2u_{x}u_{xx})
+\epsilon^{4}u_{xxxxx}=0
\eeq
having the invariants
$$
c(u)=\frac16\alpha, \quad p(u)=\frac{1-\alpha^2}{120 u}
$$
is completely integrable for $\alpha=\pm 1$ and coincides with the  second equation in the KdV hierarchy (KdVII). Varying this factor one can 
study the transition to the Kawahara equation. As expected KdVII 
shows similar oscillations as KdV \cite{gk06}, as can be seen  Fig.~\ref{kdvII4alpha}. 

For larger values of $\alpha$ one can recognize in  
Fig.~\ref{kdvII4alpha} also a formation of oscillations on the other 
side of the inflection point as in the Kawahara equation. These 
effects become smaller for larger values of $\alpha$ ($\alpha>1.2$), but it shows that the 
phenomenon of integrability with the appearance of KdV-type 
oscillations is rather subtle. Thus it seems that the decisive 
factor for the appearence and the size of these oscillations is the 
relative sign and size of the factors in front of the third and 
the fifth derivative in the equation.  Notice that  equation \eqref{kdv2} is 
for smaller $\alpha$ closer  to the non-integrable (in higher orders 
in $\epsilon$) equation 
$$
u_{t}+30u^{2}u_{x}
+\epsilon^{4}u_{xxxxx}=0
$$
(see Section \ref{sec3} above). It would be interesting to elaborate this observation in order to develop numerical tests of (approximate) integrability based on the study of the phase transition from regular to oscillatory behaviour.


\section*{Acknowledgments}
This work has been supported by the project FroM-PDE funded by the European
Research Council through the Advanced Investigator Grant Scheme. CK
thanks for financial support by the Conseil R\'egional de Bourgogne
via a FABER grant and the ANR via the program ANR-09-BLAN-0117-01.

\appendix
\section{Numerical Methods}
In this appendix we will briefly review the used methods in 
the  numerical study of the PDE in the small dispersion limit and of the 
PI2 solution and give references in which details can be found.

Since critical phenomena are generally believed to be independent on 
specific boundary conditions, we restrict our analysis to essentially 
periodic functions. Typically we consider Schwarzian functions on a 
domain on which the functions are at the boundaries smaller than machine precision 
($10^{-16}$ in double precision). Such functions can be 
periodically continued and are smooth with numerical precision. This 
allows a Fourier discretization of the spatial variables and an 
approximation of the solutions via truncated Fourier series. The use 
of Fourier spectral methods is especially efficient for the studied 
dispersive PDE because of the excellent approximation of smooth 
functions and the only minimal introduction of numerical dissipation. 
The latter is especially important if one is interested in the study 
of dispersive effects. 

After discretization of the spatial coordinates, the PDE is 
equivalent to a typically large system of ordinary differential 
equations (ODE) in the time variable. Because of the high order of the 
spatial derivatives and because of the strong gradients we want to 
study, these systems will be typically stiff. If the stiff part is 
linear as is the case for the generalized KdV equations and for the 
Kawahara equations, the system of ODE has the form
$$Lv+N[v]=0,$$
where $v$ is the discrete Fourier transform of the solution, where 
$L$ is the stiff linear operator, and where the nonlinear term $N[v]$ 
contains only derivatives of lower order. For such systems, efficient 
integration schemes exist. We use a fourth order \textit{exponential 
time differencing scheme} \cite{CM}, see \cite{etna} for a comparison 
of fourth order schemes for KdV. The numerical accuracy is controlled 
by sufficient spatial resolution, i.e., Fourier coefficients 
decreasing to at least $10^{-8}$, and by numerically checking energy 
conservation. Since all equations studied here are Hamiltonian, 
energy is a conserved quantity. Due to unavoidable numerical errors, 
it will be weakly time dependent in numerical time integrations. As 
discussed in \cite{etna}, conservation of the numerically computed 
energy typically overestimates the accuracy of a solution by two 
orders of magnitude. We always compute with an error in energy 
conservation smaller than $10^{-6}$ which implies that the error is 
well below plotting accuracy.

The situation is different for the equations with nonlinear 
dispersion in sect.~6.2. For these PDE we use an implicit fourth 
order Runge-Kutta method (Hammer and Hollingsworth method). These 
equations are numerically much more demanding. Therefore we compute 
with lower spatial resolution and an energy conservation of the order 
of $10^{-4}$.

The special solution of the PI2 equation is generated with the 
code \textit{bvp4} distributed with Matlab. For details see 
\cite{gk08}. The Hopf solution is obtained from the implicit form 
$u(x,t)=\phi(\xi)$, $x=t\phi(\xi)+\xi$ with a fixed point iteration 
to machine precision. The derivatives of the Hopf solution are 
obtained by evaluating the the analytic expressions following from the 
characteristic method.


\begin{thebibliography}{99}

 \bibitem{bmp} \'E.Br\'ezin,  E.Marinari, G.Parisi, 
 A nonperturbative ambiguity free solution of a string model. 
 {\it Phys. Lett.} {\bf B 242} (1990) 35--38.

\bibitem{bk} Yu.A.~Berezin, V.I.~Karpman, On nonlinear evolution of perturbations in plasma and other dispersive media, {\it Sov. Phys. JETP} {\bf 22} (1966) 361.
\bibitem{BI}  P. Bleher,  A. Its,  Asymptotics of the partition function of a random matrix model. {\it Ann. Inst. Fourier (Grenoble)}, {\bf 55},  (2005), no. 6, 1943-2000.
\bibitem{cg} T.~Claeys, T.~Grava, Universality of the break-up profile for the KdV equation in the small dispersion limit using the Riemann--Hilbert approach, arXiv:0801.2326,  {\it Comm. Math. Phys.} {\bf 286} (2009) 979--1009. 



\bibitem{cl1} T.~Claeys, M.~Vanlessen, The existence of a real pole-free solution of the fourth order analogue of the Painlev\'e I equation. {\it Nonlinearity} {\bf 20} (2007), no. 5, 1163--1184.

\bibitem{CM} S.~Cox and P.~Matthews, Exponential time differencing 
for stiff systems, {\it J. Comp. Phys.} \textbf{176} (2002), pp. 430-455.

\bibitem{dt} P.~Dedecker and W.M.~Tulczyjev, Spectral sequences and the inverse problem of the calculus of variations, Lecture Notes in Math. {\bf 836} (1980)
498-503.

\bibitem{dub06} B.~Dubrovin, On Hamiltonian perturbations of hyperbolic systems of conservation laws, II: universality of critical behaviour, 
{\it Comm. Math. Phys.} {\bf 267} (2006) 117 - 139.

\bibitem{nov70} B.~Dubrovin, On universality of critical behaviour in Hamiltonian PDEs, {\it Amer. Math. Soc. Transl.} {\bf 224} (2008) 59-109.

\bibitem{icmp} B.~Dubrovin,
Hamiltonian perturbations of hyperbolic PDEs: from classification results to the properties of solutions, 
In: {\it New Trends in Mathematical Physics.
Selected contributions of the XVth International Congress on Mathematical Physics}, ed. V.Sidoravicius, Springer Netherlands, 2009., pp. 231-276.

\bibitem{umn} B.~Dubrovin, S.P.~Novikov, Hydrodynamics of weakly deformed soliton lattices. Differential
geometry and Hamiltonian theory.
{\it Russian Math. Surv.} {\bf 44:6} (1989), 29-98.

\bibitem{EM}  N.Ercolani, K. McLaughlin,  Asymptotics of the partition function for random matrices via Riemann-Hilbert techniques and applications to graphical enumeration. {\it Int. Math. Res. Not.},  {\bf  14}, (2003) 755-820.

\bibitem{GX92} J. Goodman,  Zhou Ping Xin, 
Viscous limits for piecewise smooth solutions to systems of conservation laws. {\it Arch. Rational Mech. Anal. } {\bf 121} (1992), no. 3, 235-265. 

 \bibitem{gk06} T.~Grava, C.Klein, Numerical solution of the small disperion limit of the KdV equation and 
 Whitham equations, {\it Comm. Pure Appl. Math.} {\bf 60} (2007) 1623-1664.


\bibitem{gk08}T.~Grava and C.~Klein,
Numerical study of a multiscale expansion of KdV and Camassa-Holm 
equation,  in Integrable Systems and Random Matrices, ed. by 
J. Baik, T. Kriecherbauer, L.-C. Li, K.D.T-R. McLaughlin and C. 
Tomei, {\it Contemp. Math.} {\bf 458} (2008) 81-99.


\bibitem{gp} A.~Gurevich, L.~Pitaevski, Nonstationary structure of a collisionless
shock wave, {\it Sov. Phys. JETP Lett.} {\bf 38} (1974) 291--297.

\bibitem{hl} T.Y.~Hou,  P.D.~Lax, 
Dispersive approximations in fluid dynamics. 
{\it Comm. Pure Appl. Math.} {\bf 44} (1991) 1--40.

\bibitem{hunter} J.Hunter, J. Scheurle, 
Existence of perturbed solitary wave solutions to a model equation for water waves. 
{\it Phys. D}, {\bf  32} (1988), no. 2, 253-268. 

\bibitem{kapaev} A.A.~Kapaev,  
Weakly nonlinear solutions of the equation ${\rm P}\sp 2\sb 1$,
Zap. Nauchn. Sem. Leningrad. Otdel. Mat. Inst. Steklov. (LOMI) {\bf 187} (1991), {\it Differentsialnaya Geom. Gruppy Li i Mekh.} {\bf 12}, 88--109, 172--173, 175; translation in {\it J. Math. Sci.} {\bf 73} (1995), no. 4, 468--481.
\bibitem{kawa} T. Kawahara, Oscillatory solitary waves in dispersive media, {\it J. Phys. Soc. Japan},  {\bf 33} (1972),
260-264.

\bibitem{etna} C.~Klein, Fourth order time-stepping for low dispersion Korteweg-de 
Vries and nonlinear Schr\"odinger equation,  ETNA \textbf{29}, (2008) 116-135. 

\bibitem{km} Y.~Kodama, A.~Mikhailov, 
Obstacles to asymptotic integrability, 
Algebraic aspects of integrable systems, 173--204,
Progr. Nonlinear Differential Equations Appl., {\bf 26},
Birkh\"auser, Boston, MA, 1997.

\bibitem{ks} V.~Kudashev, B.~Suleimanov, A soft mechanism for the generation of dissipationless shock waves, {\it Phys. Lett.} {\bf A 221} (1996) 204--208.

\bibitem{MM} Y.~Martel and F.~Merle, Stability of Blow-Up Profile and 
Lower Bounds for Blow-Up Rate for the Critical Generalized KdV 
Equation, {\it Ann. Math.},  \textbf{155},  (2002),  235-280.


\bibitem{ponce} G. Ponce, Lax pairs and higher order models for water waves, {\it J. Differential Equations},  {\bf 102}
(1993), 360-381.

\bibitem{sul} B.~Sule\u\i manov, Onset of nondissipative shock waves and the ``nonperturbative'' quantum theory of gravitation. {\it J. Experiment. Theoret. Phys.} {\bf 78} (1994), 583--587; translated from {\it Zh. \`Eksper. Teoret. Fiz.} {\bf 105} (1994), no. 5, 1089--1097.

\bibitem{SS82}C.~Sulem, P.-L.~Sulem and H.~Frisch, Tracing	Complex	Singularities	
with	Spectral	Methods, {\it J. Comp. Phys.}	\textbf{50}, (1983) 138-161. 


\bibitem{zk} N.Zabusky, M.Kruskal, Interaction of ``solitons" in a collisionless plasma and the recurrence of initial states, {\it Phys. Rev. Lett.} {\bf 15} (1965) 2403.

\end{thebibliography}
\end{document}